\documentclass[10pt]{article}
\usepackage{amssymb,amsmath,amsthm,amsfonts}
\usepackage{pstricks,pspicture,graphicx}
\usepackage{epsfig,verbatim,alltt}
\usepackage[round]{natbib}

\def\lboxit#1{\vbox{\hrule\hbox{\vrule\kern6pt
      \vbox{\kern6pt#1\kern6pt}\kern6pt\vrule}\hrule}}

\def\thick#1{\hbox{\rlap{$#1$}\kern0.25pt\rlap{$#1$}\kern0.25pt$#1$}}

\begin{document}

\title{Issues in designing hybrid algorithms}
\author{Jeong Lee \\ School of Computing, Mathematical Sciences, \\ Auckland University of Technology, Auckland 1142, NEW ZEALAND. \\
\\ Kerrie Mengersen \\ Queensland University of Technology, Brisbane, Australia \\
\\ Christian Robert \\ Ceremade - Universit\'e Paris-Dauphine, Paris, France \\ 
\\ Ross McVinish \\ Department of Mathematics, \\ The University of Queensland, Brisbane, Australia }

\maketitle 

\begin{center}
{\sc Abstract}\vskip2mm
\end{center}

In the Bayesian community, an ongoing imperative is to develop efficient algorithms. An appealing approach is to form a hybrid algorithm by combining ideas from competing existing techniques. This paper addresses issues in designing hybrid methods by considering selected case studies: the delayed rejection algorithm, the pinball sampler, the Metropolis adjusted Langevin algorithm, and the population Monte Carlo algorithm. We observe that even if each component of a hybrid algorithm has individual strengths, they may not contribute equally or even positively when they are combined. Moreover, even if the statistical efficiency is improved, from a practical perspective there are technical issues to be considered such as applicability and computational workload. In order to optimize performance of the algorithm in real time, these issues should be taken into account. 

{\it Keywords : Delayed rejection method, importance sampling, Markov chain Monte Carlo, Metropolis-Hastings algorithm, Metropolis-adjusted Langevin algorithm, pinball sampler and population Monte Carlo}

\section{Introduction} 
Markov chain Monte Carlo (MCMC) methods, originally proposed by \citet{MetropolisRosenbluthTeller1953} and \citet{Hastings1970}, are designed to generate Markov chains with a given stationary distribution. MCMC methods have the great advantage that they apply to a broader class of multivariate problems than methods based on independent sampling. With these convenient features, MCMC methods have had a large impact on Bayesian statistics, in particular the computation of previously intractable posterior distributions. Over the last decade, a great deal of literature has been published displaying the capability of MCMC in dealing with Bayesian solutions to realistic problems in a wide variety of areas \citep{CappeRobert2000,BesagGreenHigdonMengersen1995}.\\

Another well known scheme in the Bayesian community is the importance sampling (IS) based algorithm. Instead of producing a Markov chain, this algorithm produces almost unbiased estimates using particles with associated importance weights at each iteration. These methods are extensively used to solve sequential Bayesian inference problems in which the target evolves over time \citep{LiuChen1998, DoucFreitasGordon2001}. For static problems \citet{CappeGuillinMarinRobert2004} extended IS to Population Monte Carlo (PMC) by allowing the importance function to adapt to the target in an iterative manner. This has been shown to be a computationally cheaper alternative to MCMC algorithms \citep{CelexMarinRobert2006}.\\

Over the years there has been a substantial amount of progress on the fundamental ideas of designing efficient algorithms and the theoretical properties of these methods of simulation. Each algorithm has different strengths and can be categorized by meeting different criteria based on the statistical properties of the simulated Markov chain. It is therefore natural that the question arises whether a better scheme can be developed by combining the best aspects of existing algorithms. In the following, we define the realisation of such an approach as a hybrid algorithm.\\

A hybrid algorithm can be designed from different perspectives by a variety of choices of algorithms to combine and the way in which they are combined. The primary motivation is to propose an efficient algorithm that overcomes identified weakness in the individual algorithms. It is of course important to define efficiency in this context. Existing criteria are primarily based on the statistical properties of the Markov chain \citep{AndrieuRobert2001}, but there are also computing and applicability issues to be considered in implementing algorithms for realistic problems. In this paper we consider algorithms in a more practical manner, and focus on three criteria: statistical efficiency, applicability and implementation. These are now described in more detail.

\begin{description}
\item [{\bf Statistical efficiency:}] This criterion is subject to the statistical properties of the generated Markov chain, such as the rate of convergence and the mixing speed of chains. Although classical studies ensure that the MCMC method guarantees convergence to target, this simply may not be enough to guarantee convergence in real time. The work in recent years has thus focused on theoretical developments of MCMC samplers to maximize efficiency. Established results for MCMC algorithms include geometric ergodicity for Metropolis algorithms \citep{MengersenTweedie1996, JarnerHansen2000}, and exponential ergodicity for the Langevin algorithm \citep{RobertsTweedie1996a}. When the chain is updated via an accept-reject setup, the efficiency has been characterised by deriving optimal acceptance rates \citep{RobertsTweedie1996b, RobertsGelmanGilks1997, RobertsRosenthal2001}. 

There also has been considerable attention on the mixing of a simulated Markov chain. A chain with a slower rate of mixing is more strongly dependent on the initial value and may easily become stuck in certain part of the state space for multimodal problems. This becomes a severe problem in generating reliable samples from the target as the dimension increases. Mechanisms such as the repulsive proposal \citep{MengersenRobert2003}, geometric determinant proposal \citep{MengersenRobert2003}, and delayed rejection \citep{TierneyMira1999} are proposed to improve the rate of mixing and will be examined in more detail later. 

For the IS based algorithms the convergence of estimates to their expected values is guaranteed as the number of particles increases. In practice limited numbers of particles are used and the degeneracy of importance weights is an unavoidable drawback. Attempts devoted to limit degeneracy and reduce the variance can be found in \citet{DoucGodsillAndrieu2000} among others.\\

\item [{\bf Applicability:}] In general a common feature of popular algorithms is that they are applicable in a diverse range of realistic settings. For example, the Gibbs sampler \citep{GemanGeman1984} and the Metropolis-Hastings algorithm (MHA) \citep{MetropolisRosenbluthTeller1953,Hastings1970} enjoy universal popularity because of their simple setting and in the latter case, flexible choice of the proposal distribution \citep{BesagGreenHigdonMengersen1995,BesagGreen1993,Geyer1992}. Well known hybrid approaches based on the MHA are the delayed rejection algorithm (DRA) \citep{TierneyMira1999}, the reversible jump MCMC for variable dimension models (RJMCMC) \citep{Green1995, GreenMira2001}, the Metropolis adjusted Langevin algorithm (MALA) \citep{RobertsTweedie1996a}, and the pinball sampler (PS) \citep{MengersenRobert2003}.\\

\item [{\bf Implementation:}] This criterion addresses computational issues such as the level of difficulty in coding, memory storage required and the demand on computation in real time.

In the MCMC literature, theoretical studies of statistical efficiency (first criterion above) are typically based on the number of iterations, not actual CPU time. From the perspective of practical implementation, the efficient sampling performance of an algorithm in real time (seconds) involves both the statistical efficiency and CPU time consumed in implementing the algorithm. Thus optimal performance may involve a compromise between the increase in computing workload and the gain in statistical efficiency. 

\end{description}

In this paper, we focus on selected hybrid approaches: the DRA with different types of moves for the proposal, the MALA, the PS, and the PMC, and make comparisons between the hybrid algorithms and the component algorithms on which they are based with respect to three criteria defined above. These algorithms are described in Section 2. Section 3 summarizes results of simulation studies designed to facilitate the evaluation and Section 4 describes the application of the algorithms to a real problem of analyzing particle size data. The paper concludes with a discussion in Section 5.

\section{Hybrid Approaches}

A considerable amount of attention has been devoted to the MHA which was developed by \citet{MetropolisRosenbluthTeller1953} and subsequently generalized by \citet{Hastings1970}. The advantages of the MHA are that it is straightforward to implement and is flexible with respect to the proposal density. However, it may take an unrealistically long time to explore high dimensional state spaces and tuning parameters involved with proposal density could be difficult \citep{RobertCasella2004}. In order to improve the performance of this algorithm, various forms of MHA samplers have been introduced. Before we study hybrid MCMC approaches with the MHA, we give a brief summary of the MHA itself.\\

The MHA associated with a stationary density $\pi$ and a Markov kernel
$q_1$ produces a Markov chain $\{\theta^{(t)}\}_{t=0}^T$ through the following transition.\\
%\hline
\vspace{1cm}\\
{\bf \underline{Metropolis-Hastings Algorithm (MHA)}}
\begin{description} \item [1] Generate the initial states from an initial distribution $p$, $\theta^{(0)} \sim p(\cdot)$. 
\item [2] For $t=1,\dots,T$~,
\begin{description}
\item [2.1] Generate a candidate state $\varphi$ from $\theta^{(t-1)}$ with some distribution $q_1$\\ $$\varphi\sim q_1(\theta^{(t-1)},\cdot)~.$$
\item [2.2] Calculate an acceptance probability $\alpha_1$\\
\begin{equation}\label{eq_accept} \alpha_1(\theta^{(t-1)},\varphi)=\min\left\{1,\frac{\pi(\varphi)}{\pi(\theta^{(t-1)})}\frac{q_1(\varphi,\theta^{(t-1)})}{q_1(\theta^{(t-1)},\varphi)}
\right\}~.\end{equation} \item [2.3] Accept $\varphi$ with probability $\alpha_1(\theta^{(t-1)},\varphi)$. If $\varphi$ is accepted, set $\theta^{(t)}=\varphi$. Otherwise set $\theta^{(t)}=\theta^{(t-1)}$.
\end{description} \end{description}
%\hline 

\vspace{.5cm} After a sufficient number of iterations, the initial state is forgotten and the algorithm generates samples from the target density $\pi$. The Markov chain $\{\theta^{(t)}\}^T_{t=0}$ converges to the target $\pi$ at a given rate \citep{RobertsTweedie1996b,MengersenTweedie1996}. A useful feature of the MHA is that it can be applied even when the target is only known up to a normalizing constant through the ratio $\pi/q_1$.\\

The random walk MHA in which $q_1(\varphi,\theta^{(t-1)})=q_1(\theta^{(t-1)},\varphi)$ is perhaps the simplest version of the MHA, since the probability of acceptance reduces to $\alpha_1=\min\{1,\pi(\varphi)/\pi(\theta^{(t-1)})\}$. However, the size of the proposal variance is critical to the performance of the algorithm \citep{RobertsGelmanGilks1997,RobertsRosenthal2001}. If it is too large, it is possible that the chain may remain stuck at a particular value for many iterations, while if it is too small the chain will tend to make small jumps and move inefficiently through the support of the target distribution.

\subsection{Particle system in the MCMC context} The MHA, like most MCMC algorithms to date, produces a single Markov chain simulation and monitoring focuses on its convergence of the chain to the target. An alternative particle system based on importance sampling (IS) can be used to approximate the posterior expectation of estimates of interest. The key idea of the particle system is to represent the required distribution by a set of random samples $\left(\varphi_1^{(t)}, \varphi_2^{(t)}, \cdots ,\varphi_N^{(t)} \right)$ with associated importance weights and to compute estimates based on these samples and weights.\\

In the MCMC context, instead of using the IS the whole random vector $\left(\theta_1^{(t)}, \theta_2^{(t)}, \cdots ,\theta_N^{(t)} \right)$ is resampled at each iteration according to a Markovian updating process. Once particles have eliminated their dependence on initial values, they can be considered as iid samples from the target at any given time $t$; in contrast, a long run is required in regular MCMC sampling. \citet{MengersenRobert2003} showed that an MCMC algorithm generates the production of samples of size $N$ from $\pi$ and a product distribution $\pi^{\otimes N}$ is not fundamentally different from the production of a single output from $\pi$. \\

The simplest version of a particle system in an MCMC context is a parallel run of independent MCMC algorithms. In this type of implementation the result and chains are considered individually rather than as a whole vector. The parallel system can be found in the literature on coupling methods for Markov chains \citep{BreyerRoberts2000}. \citet{MengersenRobert2003} proposed updating schemes, the repulsive proposal and the geometric determinant proposal, that allow interaction between particles. These schemes are in essence hybrid algorithms, since they combine existing algorithms, and will be examined in detail later.

\subsection{Metropolis-adjusted Langevin algorithm (MALA)} 
The random walk MHA has an advantage that it moves independently of the shape of $\pi$ and is easy to implement. However, an algorithm that is more specifically tailored to $\pi$ may converge more rapidly. This leads to the MALA that applies a Langevin diffusion approximation using the information about the target density (in the form of the derivative of $\log \pi$) to the MHA structure \citep{BesagGreen1993}. Assuming that the target distribution $\pi$ is everywhere non-zero and differentiable so that the derivative of $\log \pi$ is well defined, $\varphi$ is generated from 

\begin{equation} \label{eq Langevin 10} 
q_1(\theta^{(t-1)},\cdot)=\mathcal{N}\left (\theta^{(t-1)}+\frac{1}{2}h\nabla\log \pi(\theta^{(t-1)}), h\right )\end{equation}
where $h$ is a scaling parameter. The proposal is accepted with probability $\alpha_1(\theta^{(t-1)},\varphi)=\min\left\{1,\dfrac{\pi(\varphi)q_1(\varphi,\theta^{(t-1)})}{\pi(\theta^{(t-1)})q_1(\theta^{(t-1)},\varphi)}\right\}$. As with the traditional Euler method, scaling the step size $h$ is important. If $h$ is too large or too small, the chain may never converge to the target in real time \citep{RobertsStramer2002}. \citet{Atchade2006} showed that $h$ can be tuned adaptively.\\

The idea of shaping the candidate density based on the target was introduced as long ago as \citet{DollRosskyFriedman1978} and in the probabilistic literature has been studied by \citet{RobertsTweedie1996a,RobertsTweedie1996b}, \citet{StramerTweedie1999a,StramerTweedie1999b} and others. In discrete time space, \citet{RobertsTweedie1996a} proved that the MALA is exponentially ergodic when the tails of the target density are heavier than Gaussian. This is an advantage of MALA over the MHA that is geometrically ergodic. With a fast convergence rate the algorithm may be efficient for sampling within a single mode, but it can still fail to explore more than a single mode of the distribution. This is a major weakness of the MALA since realistic problems are often multimodal. Moreover, in practice $\nabla\log \pi$ can be expensive to compute and sometimes the first partial derivative form is not straightforward to evaluate.\\

As an attempt to cross low probability regions between modes, \citet{SkareBenthFrigessi2000} used a smoothed Langevin proposal based on a function approximation and \citet{CeleuxHurnRobert2000} demonstrated a tempering scheme using Langevin algorithms.

\subsection{Delayed rejection algorithm (DRA)} In the regular MHA, whenever a proposed candidate is rejected, the chain retains the same position so that $\theta^{(t)}=\theta^{(t-1)}$. This increases the autocorrelation along the realized path and thus increases the variance of the estimates. \citet{TierneyMira1999} proposed the DRA to ameliorate this effect. The central idea behind this algorithm is that if the candidate is rejected, a new candidate is proposed with an acceptance probability tailored to preserve the stationary distribution. Thus at the $t^{\rm th}$ iteration, if a candidate $\varphi$ is rejected with probability $\alpha_1(\theta^{(t-1)},\varphi)$, a proposal $\vartheta$ is constructed from a possibly different proposal density $q_2(\theta^{(t-1)},\varphi,\cdot)$ and $\vartheta$ is accepted with probability 
$$ \alpha_2(\theta^{(t-1)},\varphi,\vartheta)=\min \left\{1,\frac{\pi(\vartheta)q_1(\vartheta,\varphi)q_2(\vartheta,\varphi,\theta^{(t-1)})[1-\alpha_1(\vartheta,\varphi)]}
{\pi(\theta^{(t-1)})q_1(\theta^{(t-1)},\varphi)q_2(\theta^{(t-1)},\varphi,\vartheta)[1-\alpha_1(\theta^{(t-1)},\varphi)]}\right\}~.$$
The algorithm can be extended to allow proposals to be made after multiple rejections \citep{Mira2001}.\\

This scheme does not guarantee an automatic increase in the acceptance probability. However, once a better candidate is accepted, it induces an increase in the rate of acceptance and a reduction of the autocorrelation in the constructed Markov chain. \\

The drawback of the DRA is that the amount of computation and program workload increases geometrically with the number of delayed rejection attempts. The number of attempts may depend on the type of problem, model, and core questions to be answered. \citet{GreenMira2001} concluded that the workload required for three or more attempts was not worthwhile based on their limited experiments.\\

Since the proposal distributions ($q_1$ and $q_2$) are not constrained to be the same, various types of proposal distributions can be constructed using information from rejected proposals. \citet{GreenMira2001} developed a hybrid algorithm by applying the DRA to the reversible jump algorithm for trans-dimensional problems, and \citet{MengersenRobert2003} suggested the proposal scheme in a two dimensional space. We consider here the geometric deterministic proposal suggested by \citet{MengersenRobert2003} and the Langevin proposal originally suggested by \citet{MiraBressaniniMorosiTarasco2004}.

\begin{description}
\item [{\bf Geometric deterministic proposal \citep{MengersenRobert2003} :}] 
The idea is to push particles away using symmetry with respect to a line defined by the closest particle and rejected particle. Suppose that the proposal, $\varphi_i\sim q_1(\theta_i,\cdot)$ is rejected. The second proposal $\vartheta_i$ is generated by taking the reflection of $\theta_i$ with respect to the line connecting $\theta^*$ and $\varphi_i$ where $\theta^*$ is the closest particle to $\varphi_i$ among $\theta_1,\cdots \theta_{i-1}, \theta_{i+1}, \cdots,\theta_N$. This is the so-called pinball effect. \\

\item [{\bf Langevin proposal :}] If the current proposal $\varphi$ is rejected, the second proposal $\vartheta$ is generated according to 
$$q_2(\theta^{(t-1)},\varphi,\cdot)=\mathcal{N}\left (\varphi+\frac{1}{2}h\nabla\log \pi(\varphi), h\right ) ~.$$ 
\end{description}

\subsection{Pinball Sampler (PS)} \citet{MengersenRobert2003} introduced a hybrid algorithm, the PS, which is a combination of the MHA, delayed rejection mechanism, particle system, and repulsive proposal. It is an updating system for a particle system that is based on a standard random walk with correction to avoid the immediate vicinity of other particles.\\

Unlike traditional particle systems, neither importance sampling schemes nor weights are used. At each iteration, the whole vector $\left(\theta_1^{(t)},\theta_2^{(t)},\dots,\theta_{N}^{(t)}\right)$ is updated. Thus the algorithm produces iid samples from the target, which is an advantage over MCMC methods. Moreover, at each vector update the repulsive mechanism discourages particles from being too close and hence avoids possible degeneracy problems.\\

We now study the properties of the repulsive proposal density in order to understand the mechanics of the PS.

\begin{description}
\item [ {\bf Repulsive proposal (RP)}] 

The RP suggested by \citet{MengersenRobert2003} uses the following pseudo distribution,
\begin{equation} \label{eq PS 10}
\pi^R(\theta_i)\propto \pi(\theta_i)\prod_{j\neq i}e^{-\xi/\pi(\theta_j)\parallel \theta_i-\theta_j\parallel^2}\end{equation}
where $\xi$ is a tempering factor.

The pseudo distribution, $\pi^R$, is derived from the distribution of interest $\pi$ by multiplying exponential terms that create holes around the other particles $\theta_j$, $(i\neq j)$. These thus induce a repulsive effect around the other particles. The factor $\pi(\theta_j)$ in the exponential moderates the repulsive effect in zones of high probability and enhances it in zones of low probability. If the tempering factor, $\xi$, is large, $\pi^R$ is dominated by the repulsive term and becomes very different from the target distribution $\pi$. If $\xi$ is small, there is a negligible repulsive effect on $\pi^R$, so $\pi^R\approx \pi$.\\

The MHA is easily adapted to include the repulsive proposal through two acceptance steps, firstly with $\pi^R$ and secondly with the true posterior $\pi$ to ensure convergence to the target distribution. The repulsive proposal algorithm is as follows.\\
%\hline 

\vspace{.5cm} {\bf \underline {Metropolis Hastings algorithm with the repulsive proposal}}
\begin{description} 
\item [1] Generate initial particles from an initial distribution $p$, $\theta_{1,\dots,N}^{(0)}\sim p(\cdot)$.
\item [2] For $t=1,\dots,T$, \\ For $i=1,\dots,N$,
\begin{description}
\item [2.1] Generate $\varphi_i\sim q_1(\theta^{(t-1)}_i,\cdot)$.
\item [2.2] \underline{\normalsize \it Propose Step} \\ \vspace{-.3cm}\\ 
Determine the probability using $\pi^R$,
$$\alpha_1^*(\theta^{(t-1)}_i,\varphi_i)=\min\left\{1,\frac{\pi^R(\varphi_i)}{\pi^R(\theta^{(t-1)}_i)}\frac{q_1(\varphi_i,\theta^{(t-1)}_i)}{q_1(\theta^{(t-1)}_i,\varphi_i)}\right\}~.$$
Generate $r\sim U[0,1]$. If $r<\alpha_1^*$, go to {\it Correction Step}. Otherwise set $\theta^{(t)}_i=\theta^{(t-1)}_i$.
\item [2.3] \underline{\normalsize \it Correction Step}\\ \vspace{-.3cm}\\
Implement a final Metropolis acceptance probability calibrated for the target distribution $\pi$,
$$\alpha_1(\theta^{(t-1)}_i,\varphi_i)=\min\left\{1,
\frac{\pi(\varphi_i)}{\pi(\theta^{(t-1)}_i)}\frac{\pi^R(\theta^{(t-1)}_i)}
{\pi^R(\varphi_i)} \right \}~.$$
Accept $\varphi_i$, with probability $\alpha_1$. If $\varphi_i$ is accepted, set $\theta_i^{(t)}=\varphi_i$; otherwise set $\theta_i^{(t)}=\theta^{(t-1)}_i$.
\end{description} \end{description}
%\hline

\vspace{1cm} The tempering factor $\xi$ can be calibrated during the simulation against the number of particles $N$ and the acceptance rate. The value should be tuned such that proposals are easily rejected if they are generated from low probability regions or are too close to existing particles.
\end{description}

Borrowing the idea from the DRA, two or more attempts at updating may be pursued using the information from rejected proposals, and the delayed rejection step can easily be included in this algorithm.\\

The two-stage PS algorithm for generating from a target distribution $\pi$ as follows.\\
%\hline

\vspace{.5cm} {\bf \underline{Pinball Sampler (PS)}}
\begin{description}
\item [1] Generate initial particles from an initial distribution $p$, $\theta_{1,\dots,N}^{(0)}\sim p(\cdot)$.
\item [2] For $t=1,\dots,T$,\\ For $i=1,\cdots,N$,
\begin{description}
\item [2.1] \underline{\it First stage} \begin{description}
\item [2.1.1] Generate $\varphi_i\sim q_1(\theta_i^{(t-1)},\cdot)$.
\item [2.1.2] Generate $r_1\sim U[0,1]$. If $r_1 < \alpha_1^*(\theta_i^{(t-1)},\varphi_i)$, accept $\varphi_i$ as in the {\it Correction Step}. Either $r_1 > \alpha_1^*(\theta_i^{(t-1)},\varphi_i)$ or $\varphi_i$ is rejected with $\pi$, go to the {\it Second stage}. \end{description}
\item [2.2] \underline{\it Second stage} \begin{description}
\item [2.2.1] Draw a candidate $\vartheta_i \sim q_2(\theta^{(t-1)}_i,\varphi_i,\cdot)$. Here the $q_2$ is the geometric deterministic proposal.
\item [2.2.2] Generate $r_2\sim U[0,1]$. If $r_2 < \alpha_2^*(\theta_i^{(t-1)},\varphi_i,\vartheta_i)$, accept $\vartheta_i$ as in the {\it Correction Step}.
\end{description}
\end{description} \end{description}
%\hline

\vspace{.5cm} Note that other proposals can be considered for $q_2$ for different problems. In particular, some stochasticity could augment the deterministic proposal described above.

\subsection{Population Monte Carlo algorithm (PMC)} 
The PMC algorithm by \citet{CappeGuillinMarinRobert2004} is an iterated IS scheme. The major difference between the PMC algorithm and earlier work on particle systems is that the PMC algorithm uses a resample step to draw samples from a target that does not evolve with time.\\ 
%\hline

\vspace{.5cm} {\bf \underline {Population Monte Carlo (PMC) algorithm}}
\begin{description} \item [1] {\it Initialization,}\\ Generate initial particles from an initial distribution $p$, $\theta^{(0)}_{1,\dots,N}\sim p(\cdot)$.\\
\item [2] For $t=1,\dots,T$,
\begin{description} 
\item [2.1] Construct the importance function $g^{(t)}$ and sample $\varphi_1^{(t)},\cdots, \varphi_N^{(t)} \sim g^{(t)}(\cdot)$. 
\item [2.2] Compute the importance weight
$$\omega^{(t)}_i\propto \frac{\pi\left (\varphi^{(t)}_i \right )}{g^{(t)}\left (\varphi^{(t)}_i\right )}~.$$
\item [2.3] Normalize the importance weights to sum to 1.
\item [2.4] Resample with replacement $N$ particles $\{\theta^{(t)}_i;i=1,\dots,N\}$ from the set $\{\varphi^{(t)}_i;i=1,\dots,N\}$ according to the normalized importance weights.
\end{description} \end{description}
%\hline

\vspace{.5cm} The advantage of the PMC over the MCMC algorithms is that it produces (approximately) unbiased samples at each iteration and the adaptive perspective can be achieved easily. \citet{DoucGuillinMarinRobert2007} proposed a $D$-kernel PMC that automatically tunes the scaling parameters and induces a reduction in the asymptotic variance via a mixture of importance functions.\\

Resampled particles according to the importance weights are relatively more informative than those that are not resampled and this available information can be used to construct the $g$ at the next iteration. A common choice for $g$ is a set of individual normal distributions for $\theta_1,\dots, \theta_N$ so that the importance function of particle-$i$ is defined as a normal distribution centered at $\theta_i$ \citep{CappeGuillinMarinRobert2004,DoucGuillinMarinRobert2007}.\\

In this paper we consider a different approach such that $g$ is based on the population of $\theta_1,\dots, \theta_N$ and its evaluation of particle-$i$ involves all resampled particles, not just $\theta_i$. The nonparametric kernel estimate of a density of $\theta$ is given by fitting a mixture of $N$ components in equal proportions $1/N$,
$$\hat{\pi}(\theta) = \dfrac{1}{N\vartheta} \sum^N_{i=1} \mathcal{K}\left( \dfrac{\theta'-\theta_i}{\vartheta}\right)~,$$

where $\mathcal{K}$ is some kernel. When $\mathcal{K}$ is a standard Gaussian function with mean zero and variance 1, the equation simplifies to a mixture of Gaussian functions with mean $\theta$ and variance $x^2$ 
\begin{equation} \label{eq PMC 50}
g(\theta') = \dfrac{1}{N} \sum^N_{i=1} \mathcal{N} (\theta'-\theta_i,x^2)~.\end{equation}

The main interest in the kernel estimate of density literature is the choice of optimal values for $\theta'$ so that the density is neither undersmoothed nor oversmoothed \citep{Scott1992}. Since the choice of $g$ with tails that are fatter than $\pi$ guarantees the existence of the variance of the estimate from a theoretical point of view, we multiply the optimal value suggested by \citet{Scott1992} by $k>1$, so that 

$$x = k \sqrt{\sigma^2_{\theta}} N^{-1/(D+4)}$$
where $\sigma^2_{\theta}$ denotes the variance of $\theta_1,\dots, \theta_N$ and $D$ is the dimension of $\pi$.\\

As with other IS based algorithms, the downside of the PMC algorithm is that degeneracy may occur due to instability of the importance weight for the multimodal target as the dimension increases. The degeneracy is observed as the importance weight is concentrated on very few particles or a single particle. Theoretically this phenomena can be avoided when the number of particles increases exponentially with the dimension \citep{LiBengtssonBickel2005} but this usually causes storage problems.\\

As an attempt to improve this degeneracy phenomena we now adapt the repulsive effect described in Section 2.4.1 to the PMC algorithm.

\subsubsection{PMC with the repulsive effect} 
The idea of this hybrid approach is to generate particles from the importance function with the repulsive effect in the standard PMC context. Due to the repulsive effect proposed particles are not too close to each other and are still in a relatively informative region via a kernel density estimate based on resampled particles from the last iteration. Let $\varphi_1,\dots,\varphi_N$ be random variables almost from $g$ as in (\ref{eq PMC 50}). The pseudo distribution with the repulsive effect is

\begin{equation} \hat{g}(\varphi_i) \propto g(\varphi_i) \left((1-\nu)+\nu \prod^N_{j\neq i}e^{-\xi/g(\varphi_j)\|\varphi_i-\varphi_j\|^2}\right)~. \end{equation}
 
This form consists of one more parameter than the $\pi^{R}$ used in the MCMC algorithm and the size of hole is adjusted by $\xi$ and $\nu$; the $\xi$ indicates the width of holes, and the $\nu$, ($0<\nu <1$) indicates the depth of holes. If both $\nu$ and $\xi$ are very small, negligible holes are created around particles, and proposals are almost from the $g$. For large $\nu$ and $\xi$, particles are pushed away with greater repulsive effect due to bigger holes. In particular if $\xi$ is very large with respect to the number of holes $N$, the holes overlap with each other and $\hat{g}$ is oversmoothed. These hyperparameters need to be tuned with respect to the total number of holes and a scaling parameter $k$.\\

In order to apply $\hat{g}$ to the IS successfully, its normalizing constant should be known. Since it is not solvable analytically, we approximate the normalizing constant, $C$,
\begin{eqnarray}  
\dfrac{{\displaystyle \int}\hat{g}(\varphi) ~d\varphi }{{\displaystyle \int} g(\varphi)~d\varphi} &=& \dfrac{{\displaystyle \int}{ \dfrac{\hat{g}(\varphi)}{g(\varphi)-\hat{g}(\varphi)}\dfrac{g(z)-\hat{g}(\varphi)}{C} ~d\varphi }}{{\displaystyle \int} \dfrac{g(\varphi)}{g(\varphi)-\hat{g}(\varphi)}\dfrac{g(\varphi)-\hat{g}(\varphi)}{C} ~dz }~=~C \nonumber \\
& \approx&\dfrac{\dfrac{1}{M}{\displaystyle \sum^M_{m=1}} \dfrac{\hat{g}(\varphi_m)}{g(\varphi_m)-\hat{g}(\varphi_m)} }{\dfrac{1}{M}{\displaystyle \sum^M_{m=1}} \dfrac{g(\varphi_m)}{g(\varphi_m)-\hat{g}(\varphi_m)} }~=~C' \label{NomCont}
\end{eqnarray}
where $p(\varphi_1,\dots, \varphi_M) \propto g-\hat{g}$ and $M\geq N$. For reasonable approximation of $C$, adjusting $M$ with respect to the width and the depth of holes is necessary.\\

%\hline

\vspace{.5cm} {\bf \underline {PMC with the repulsive effect}}
\begin{description} \item [1] {\it Initialization,}\\ Generate initial particles from an initial distribution $p$, $\theta^{(0)}_{1,\dots,N}\sim p(\cdot)$.\\
\item [2] For $t=1,\dots,T$,
\begin{description}
\item [2.1] Construct the $g^{(t)}$ based on $\theta_1^{(t-1)},\dots,\theta_N^{(t-1)}$ and sample $\tilde{\varphi}_1^{(t)},\cdots, \tilde{\varphi}_N^{(t)}$ from $g^{(t)}$. 
\item [2.2] Construct the $\hat{g}^{(t)}$ by creating holes at $\tilde{\varphi}_1^{(t)},\cdots, \tilde{\varphi}_N^{(t)}$. Generate $\tilde{\varphi}_{N+1}^{(t)},\cdots, \tilde{\varphi}_M^{(t)}$ from $g^{(t)}$ that satisfy $g^{(t)}(\tilde{\varphi})>\hat{g}^{(t)}(\tilde{\varphi})$. Estimate the $C'$ based on $\tilde{\varphi}_1,\dots, \tilde{\varphi}_M$.
\item [2.3] Generate $\varphi_1^{(t)},\cdots, \varphi_N^{(t)}$ from $\hat{g}^{(t)}$.
\item [2.4] Compute the importance weight
$$\omega^{(t)}_i\propto C' \frac{\pi\left (\varphi^{(t)}_i \right )}{\hat{g}^{(t)}\left (\varphi^{(t)}_i\right )}~.$$
\item [2.5] Normalize the importance weights to sum to 1.
\item [2.6] Resample with replacement $N$ particles $\{\theta^{(t)}_i;i=1,\dots,N\}$ from the set $\{\varphi^{(t)}_i;i=1,\dots,N\}$ according to the normalized importance weights.
\end{description} \end{description}
%\hline

\vspace{1cm} In this approach samples $\tilde{\varphi}_1,\dots,\tilde{\varphi}_M$ are generated to create holes and estimate $C'$. Through this process proposals $\varphi_1,\dots, \varphi_N$ can be generated from $\hat{g}$. As will be shown later, when $\xi$ is small, simply $M=N$ gives a reasonable approximation and no extra particles are required.\\

Both PMC and its hybrid approach produce estimates of the posterior expected values using IS. To assess the characteristics of the algorithm over iterations, the simulation is summarized by a series of estimates, 
$$\left(\sum^N_{i=1}\varphi^{(1)}_i \bar{w}^{(1)}_i~, \dots, \sum^N_{i=1}\varphi^{(T)}_i \bar{w}^{(T)}_i\right)~.$$

\section{Illustration of hybrid algorithms} 
In this section we examine the hybrid methods described in Section 2 and study how individual components of the algorithms influence the properties of the overall algorithm. Algorithms are programed using version 7.0.4 MATLAB software and run by the Lyra that is an SGI Altix XE Cluster containing $112 \times 64$ bit Intel Xeon cores. The investigation is first carried via simulation of the hybrid MCMC algorithms using a toy example in Section 3.1 and 3.2, and the sensitivity of a tempering factor of the RP is given in Section 3.3. Section 3.4 presents a simulation study of the PMC and its hybrid approach.

\subsection{Simulation study 1}

We commence our investigation by choosing a target distribution that comprises two well separated modes, described by a mixture of two dimensional normal distributions
$$\pi=\dfrac{1}{2} \mathcal{N}([0,0]^T,I_2) +\dfrac{1}{2} \mathcal{N}([5,5]^T,I_2)~. $$
The expectations of the two components are $\mathbb{E}_{\pi}(\theta_1)=\mathbb{E}_{\pi}(\theta_2)=2.5$. A normal proposal distribution $q_1(\theta,\cdot)=\mathcal{N}(\theta,sI_2)$, $s>0$ is used for the MHA, the RP, and as the first-stage proposal for the DRA and the PS.\\

We consider two simulation studies. The first study focuses on the performance of the algorithms in terms of statistical efficiency, and relative cost of computation. The second study focuses on the mobility of the chains in a special setup where initial values are generated from a certain part of state space and demonstrates the ability of the algorithm to detect the two modes of the target distribution.

\subsubsection{Performance study}
In this paper the comparison of the performance of algorithms is based on 100 replicated simulations of each algorithm. Each simulation result is obtained after running the algorithm for 1200 seconds using 10 particles. This allows a comparison of the algorithms in a more realistic way by fixing the running time and observing the consistency of estimates throughout the replications.\\

For each simulation, the performance is defined as the accuracy of estimating $H$, $\mathbb{E}_{\pi}[H(\theta_1,\theta_2)] = \mathbb{E}_{\pi}(\theta_1)$ and the variance ($\sigma^2_H$), the efficiency of moves indicated by the rate of acceptance ($A$), the computational demand indicated by the total number of iterations ($T$), the mixing speed of the chain estimated by the integrated autocorrelation time ($\tau_{\pi}(H)$), and the loss in efficiency due to the correlation along the chain by the effective sample size ($ESS_H$) where $ESS_H=(T-T_0)/\tau_{\pi}(H)$ and $T_0$ is the length of the burn-in. The term $\tau_{\pi}$ indicates the number of correlated samples with the same variance-reducing power as one independent sample, and is a measure of the autocorrelation in the chain. The $ESS_H$ can be interpreted as the number of equivalent iid samples from the target distribution for 1200 seconds of running and provides a practical measure for the comparison of algorithms in terms of both statistical and computational efficiency. Except for $T$ and $A$, all measurements are estimated after ignoring the first 500 iterations ($T_0=500$).\\

Based on the 100 replicates, these measures are summarized by taking the respective averages ($\bar{H}$, $\bar{\sigma}^2_H$, $\bar{A}$, $\bar{T}$, $\bar{\tau}_{\pi}(H)$, and $\overline{ESS}_H$) and mean square error (MSE). The MSE measures the consistency of performances over 100 replicates and the stability of the algorithm. For instance a small MSE can be interpreted to indicate that the algorithm will produce a reliable output with a given expected statistical efficiency. The performance of the single and parallel chain implementation of the MHA, the MHA with the RP, the DRA using three moves for the second-step proposal and the MALA using a traditional proposal and smoothed Langevin proposal are summarized in Table \ref{table 1} and Table \ref{table 2} for two different sizes of scaling parameters, $s=4,2$ and $h=4,2$, respectively. All chains converged to the target reasonably well based on informal diagnostics. The performance of each algorithm is discussed below.

\begin{table} 
\centering \begin{tabular} {c|c c c c c c}
 & $\overline{T}$ & $\bar{A}$ & $\bar{H}$ & $\bar{\sigma}^2_H$ & $\bar{\tau}_{\pi}(H)$ & $\overline{ESS}_H$ \\ \hline 
{\small True value} & & & 2.5 &&& \\ \hline
$MHA^s$ & $1.0376\times 10^6$ & 0.30 & 2.4979 & 7.2523 & 294.7127 & 3507.2 \\
&(456.0547)&($<10^{-4}$)&(0,0032)&(0.0013)&(0.9197)&(10.8990) \\
$MHA$ & $2.4259\times 10^6$ & 0.30 & 2.5002 & 0.7250 & 296.5254 & 2181.7 \\
&($1.2\times 10^4$)&($<10^{-4}$)&(0.0007)&(0.0005)&(0.5736)&(43.1207) \\ \hline
$MHA^{RP}$ & $1.2172\times 10^5$ & 0.29 & 2.4904 & 0.7213 & 266.5477 & 371.2440 \\
($\xi = 10^{-5}$) &(581.89)&($<10^{-4}$)&(0.0036)&(0.0028)&(2.4224)&(3.8696) \\ \hline
$DRA$ & $1.0012\times 10^6$ & 0.49 & 2.4978 & 0.7187 & 196.8483 & 5086.9 \\
&(24.8503)&($<10^{-4}$)&(0.0009)&(0.0006)&(0.5143)&(13.1896) \\
$DRA^{LP}$ & $1.9914\times 10^5$ & 0.34 & 2.5010 & 0.7223 & 253.4174 & 786.2 \\
($h$=4) &(1031.5)&($<10^{-4}$)&(0.0020)&(0.0018)&(1.3601)&(6.0554) \\ 
$DRA^{Pinball}$ & $6.4729\times 10^5$ & 0.61 & 2.5025 & 0.7235 & 239.1775 & 2708.4 \\
&(3091.8)&($<10^{-4}$)&(0.0010)&(0.0009)&(0.8232)&(17.3921) \\ \hline
$MALA$ & $2.0415\times 10^5$ & 0.29 & 2.4978 & 0.7265 & 828.4097 & 505.0 \\
($h=4$) &(241.9826)&($<10^{-4}$)&(0.0093)&(0.0068)&(56.5102)&(60.2510) \\ 
\hline
\end{tabular} \caption{Comparing performance in the two dimensional example with the proposal variance $s=4$. The values in the bracket are the MSE of measures.} \label{table 1} \end{table}

\begin{table} \centering \begin{tabular} {c|c c c c c c}
 & $\overline{T}$ & $\bar{A}$ & $\bar{H}$ & $\bar{\sigma}^2_H$ & $\bar{\tau}_{\pi}(H)$ & $\overline{ESS}_H$ \\ \hline 
{\small True value} & & & 2.5 &&& \\ \hline
$MHA^s$ & $1.037\times 10^6$ & 0.43 & 2.4999 & 7.2473 & 770.6577 & 1339.7 \\
&(346.7434)&($<10^{-4}$)&(0.0053)&(0.0014)&(1.7493)&(3.0647) \\
$MHA$ & $2.481\times 10^6$ & 0.43 & 2.4992 & 0.7253 & 872.3864 & 2845.7 \\
&($2.52\times 10^4$)&($<10^{-4}$)&(0.0012)&(0.0009)&(2.5346)&(29.9846) \\ \hline
$MHA^{RP}$ & $1.2294\times 10^5$ & 0.42 & 2.4952 & 0.7230 & 786.3059 & 157.7317 \\
($\xi = 10^{-5}$) &($1.37\times 10^3$)&(0.0016)&(0.0062)&(0.0044)&(9.7029)&(2.3711) \\ \hline 
$DRA$ & $1.0012\times 10^5$ & 0.63 & 2.4983 & 0.7148 & 700.9815 & 1429.6 \\
&(41.2523)&($<10^{-4}$)&(0.0015)&(0.0013)&(2.6121)&(5.2876) \\
$DRA^{LP}$ & $2.2856\times 10^5$ & 0.61 &  2.5021 & 0.7235 & 553.5437 & 413.6313 \\
($h=2$) &($1.77\times 10^3$)&($<10^{-4}$)&(0.0031)&(0.0024)&(3.5946)&(4.0817) \\ 
$DRA^{Pinball}$ & $6.8763\times 10^5$ & 0.68 & 2.4990 & 0.7218 & 674.7575 & 1020.0 \\
&($5.21\times 10^3$)&($<10^{-4}$)&(0.0021)&(0.0015)&(3.0249)&(8.6733) \\
\hline
$MALA$ & $41.9834\times 10^5$ & 0.67 & 2.5069 & 0.6916 & 38.0650 & 5839.6 \\
($h=2$) &($1.66\times 10^3$)&($<10^{-4}$)&(0.0166)&(0.0122)&(1.3820)&(198.9234) \\
\hline
\end{tabular} \caption{Comparing performance in the two dimensional example with the proposal variance $s=2$. The values in the bracket are the MSE of measures.} \label{table 2} \end{table}

\begin{description}
\item [{\bf MHA}] It is noticeable that due to the parallel computing the $MHA$ takes the shortest CPU time per iteration and produces the largest sample size, 20 times more proposals than the single chain, $MHA^s$. The improvement in bias of samples from the parallel chains is not observable compared to a single chain. To optimize the performance the proposal variance $s$ is usually tuned such that $A\approx 0.3$ \citep{RobertsRosenthal2001} and for this example $s=4$ is the optimal value. \\

\item [{\bf MALA}] When $h$ is not sufficiently large it is seen that samples are easily trapped in a certain part of the state space. It is known that often the $MALA$ performs poorly for multimodal problems, as the chain is pulled back toward the nearest mode and may become stuck \citep{SkareBenthFrigessi2000}. This is observed by a larger MSE of measures compared with other MCMC algorithms in general. In particular when $s=2$ a very small $\tau_f$ induces a huge ESS and corresponding MSE of $H$. \\

\item [{\bf MHA with RP ($MHA^{RP}$)}] It is seen that the repulsive effect induces a fast mixing chain. The repulsive proposal reduces the autocorrelation along the chain by pushing particles apart around a neighborhood. Its effect is more obvious when the proposal variance is relatively small.

The repulsive proposal imposes an additional tuning parameter $\xi$ on the $MHA^{RP}$ compared to the $MHA$. The algorithm is illustrated with $\xi=10^{-4}$, and approximately 99.5\% of proposals that are accepted with $\pi^R$ are accepted with $\pi$. The choice of $\xi$ is critical in implementing the RP and we will discusss this matter later. 

The drawback of the $MHA^{RP}$ is that it is expensive to compute. For $D$ dimensional problems with $N$ particles and an arbitrary $N\times D$ matrix $\theta^{(t)}$ the repulsive part of a pseudo distribution adds $O(ND)$ operations.

This algorithm also can be unstable and as for the MALA, the MSE of measures tends to be large in some circumstances. As illustration, given with $\xi$ carefully tuned, we observed six exceptional measures out of 100 replicates using $s=2,4$; see Table \ref{table exc val}. A very low rate of acceptance, a high $\tau_{\pi}$, and an extremely small ESS indicate that the chain is stuck in a certain region and does not adequately explore the state space. However these instances occurred relatively rarely, and the most of times the algorithm performs reasonably well. 

\begin{table} \begin{center} \begin{tabular} {c| c c c c c c}
\hline
 $s$& $T$ & $A$ & $H$ & $\sigma^2_H$ & $\tau_{\pi}(H)$ & $ESS_H$ \\ \hline 
 2 &118283 & 0.05 & 3.2699 & 0.2863 & 3641 & 32.32\\
  & 114805 & 0.06 & 3.4432 & 0.6235 & 5907 & 19.35\\ \hline
4 &122372 & 0.0009 & 1.4894 & 0.4307 & 10887 & 11.19\\ 
 & 122107 & 0.0461 & 3.1459 & 0.5232 & 1817 & 66.94\\
 & 124531 & 0.1922 & 2.9552 & 1.1092 & 602 & 206.17 \\
 & 124411 & 0.1878 & 1.9514 &  0.5510 & 423 & 292.69\\ \hline
\end{tabular} \caption{Exceptional estimates of $MHA^{RP}$.} \end{center} \label{table exc val} \end{table}

\item [{\bf DRA}]
By making the second attempt to move instead of remaining in the current state, algorithms with a delayed rejection mechanism produce less correlated samples and have a higher rate of acceptance. The demand in computation is also increased and differs with different types of moves. We examined three types of moves for the second-step proposal, the normal random walk as for the first proposal ($DRA$), the geometric deterministic proposal ($DRA^{Pinball}$) and the Langevin proposal ($DRA^{LP}$). 

With the geometric deterministic proposals, over 60\% of proposed moves were accepted regardless of the value of the proposal variance. The geometric deterministic proposal pushes away from the closest particle; however with a moderate repulsive probability proposed particles may remain in the neighborhood of existing particles, in which case it is highly likely to be accepted.

Generally the Langevin proposal seems a suitable choice for the second-stage proposal in terms of the improvement in the mixing speed of chains when the proposal variance is not relatively large. However it requires a greater amount of computation than other types of proposal, and despite a diminished $\tau_{\pi}$ the ESS is relatively small. In this case the loss in the computational efficiency overwhelms the gain in the statistical efficiency. Given the nature of the Langevin diffusion tuning of $h$ is essential.

The normal random walk for a second stage proposal is faster to compute and results in improved mixing with a sufficiently flexible choice of the proposal variance. As seen in Table \ref{table 1}, when particles can move around the state space efficiently, neither the deterministic proposal nor the Langevin diffusion approximation will induce a substantive reduction in the correlation of the samples. 

\end{description}

\subsubsection{Ability to detect modes}
We now consider a burn-in period in which chains are still dependent on initial states. A strong dependence of chains on the starting value is a well known problem with the MHA. We examine the selected hybrid approaches to study how quickly chains move from initial values and find other local modes.\\

We run algorithms in the same manner with initial particles from one particular mode $\mathcal{N}([0,0], I_2)$, $s=2$ for the random walk, and $h=2$ for the Langevin proposal. Through 400 repeated simulations, the ability to detect the other mode $\mathcal{N}([5,5], I_2)$ is estimated. The allocation of particles is determined by the distance from the center of mode.\\

\begin{figure}[ht!] \begin{center} \setlength{\unitlength}{1cm} \begin{picture}(12,5) 
\put(-1,0){\includegraphics[width=6.5cm,height=5cm]{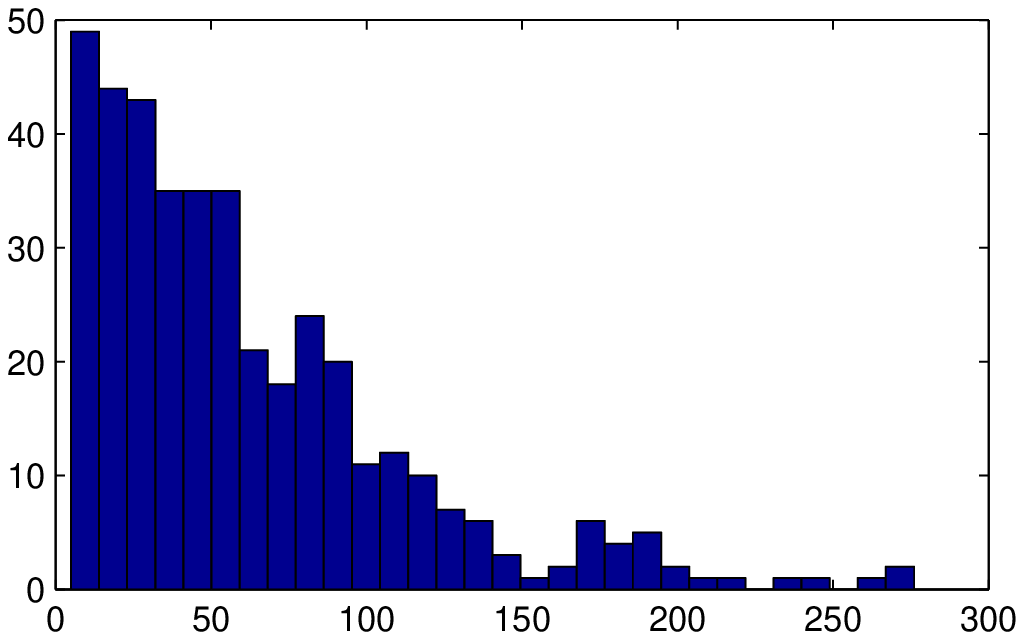}} 
\put(6,0){\includegraphics[width=6.5cm,height=5cm]{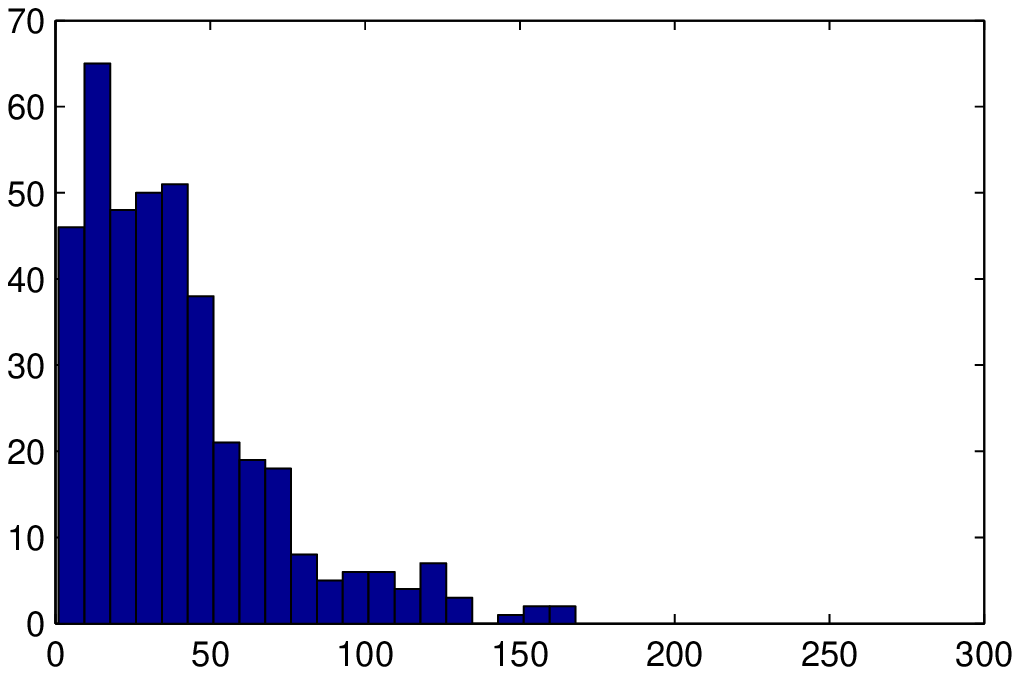}}
\rput(4.5,4.3){a)} \rput(11.5,4.3){b)}
\end{picture} \end{center} \caption{Histogram of the number of iterations before the second mode is discovered via running a parallel chain of (a) the MHA, and (b) the DRA with the Langevin proposal over 400 replicates.}\label{figure 4}\end{figure}

Figure \ref{figure 4} shows a histogram of the number of iterations before the second mode is detected for the first time during 400 replicates of a parallel running of the MHA, and the DRA with the Langevin proposal. Generally particles escaped from the first mode relatively quickly but there is a noticeable number of chains for which this was not the case. The chains with a slow mixing rate ($MHA$) tended to have a longer tail than the faster chain ($DRA^{LP}$).\\

Table \ref{table 3} shows the total number of trials in which the second mode was detected within the first 50 iterations during 400 repeats. Approximately 86.5\% of the MALA simulations failed to detect both modes. Generally the DRA had a better ability to detect modes than the MHA. In particular 72.5\% of the $DRA^{LP}$  trials found the second mode within 50 iterations, compared to only 50.8\% for the MHA.

\begin{table} \centering 
\begin{tabular} {l |c || l| c } \hline
MHA & 203 & $MHA^{RP}$ & 211 \\
DRA & 252 & $DRA^{Pinball}$ & 250\\
$DRA^{LP}$ & 290 & MALA & 54\\ \hline
\end{tabular} \caption{The total number of cases in which the second mode is detected within the first 50 iterations during 400 replicates.} \label{table 3} \end{table}

\subsection {Simulation study : PS}

The PS is interesting to study as a hybrid algorithm because it combines features of four schemes, the MH, DRA, RP, and particle system. We study the relative contribution of these individual algorithms and their combined character through a simulation study analogous to that described in Sections 3.1.1 and 3.1.2.\\

The performance study of the PS is carried out in the same manner to Section 3.1.1, and a summary of results is given in Table \ref{table PS}. Comparing the results of the PS to component algorithms in Tables \ref{table 1} and \ref{table 2}, the PS produces less correlated samples than the MHA and has a high rate of acceptance due to the DRA and pinball effect. A previous simulation study in Section 3.1 has shown that the repulsive proposal and the geometric deterministic proposal are better choices than the small random walk proposal; however when the size of random walk is sufficiently large, the normal random walk proposal is better. This was apparent in this simulation study. In particular when $s=2$, $\bar{\tau}_{\pi}$ was smaller and $\bar{A}$ was greater than for the individual algorithms, a single and parallel chain of the MHA, MHA with RP, and the DRA. This was supported by the results of the mode detecting test in which 245 out of the 400 repeats found the second mode within 50 iterations. It seems that overall the statistical efficiency of the PS is most affected by the DRA.\\

\begin{table} \begin{center} \begin{tabular} {c| c c c c c c c}
 & $s$ & $\overline{T}$ & $\bar{A}$ & $\bar{H}$ & $\bar{\sigma}^2_H$ & $\bar{\tau}_{\pi}(H)$ & $\overline{ESS}_H$ \\ \hline 
{\small True value} & & & & 2.5 &&& \\ \hline 
$PS$ & 4 & 95226 & 0.56 & 2.4904 & 0.7213 & 266.5477 & 371.2440 \\
($\xi=10^{-5}$) &&(891.984)&(0.0067)&(0.0268)&(0.0104)&(6.8292)&(8.2292) \\
 & 2 & 93550 & 0.64 & 2.4971 & 0.7148 & 679.0638 & 141.1047 \\
 &&(972.6976)&(0.0061)&(0.0240)&(0.0074)&(11.1321)&(2.6741) \\
\hline
\end{tabular}\end{center} \caption {Performance of the PS.} \label{table PS} \end{table}
\vspace{1cm}

As seen for the MHA with the RP, some instability of performance and exceptional values of $ESS_H$ are seen with the PS. This phenomenon is illustrated in Table \ref{table exc ps}.

\begin{table} \begin{center} \begin{tabular} {c| c c c c c c}
\hline $s$& $T $& $A$ & $H$ & $\sigma^2_H$ & $\tau_{\pi}(H)$ & $ESS_H $ \\ \hline 
2 & 82483 & 0.0006 & 0.8945 & 0.1475 & 10035 & 8.17 \\ 
 & 82931 & 0.0509 & 3.1990 & 0.4323 & 1999 & 41.24 \\
 & 75580 & 0.0269 & 3.7948 & 0.2841 & 2666 & 28.16 \\  \hline
4 & 100708 & 0.0034 & 1.8624 & 0.3980 & 8512 & 11.77 \\ 
& 74653 & 0.0594 & 1.6395 & 0.4539 & 668 & 110.98 \\ \hline
\end{tabular} \end{center} \caption{Exceptional estimates of the PS.} \label{table exc ps} \end{table} 

The repulsive proposal adds a tuning parameter $\xi$ and the geometric deterministic form of the proposal considered here gives a restriction on the size of dimension. These issues thus impacted on the implementation of the PS. The major drawback of the PS was a large workload in terms of computation due to the repulsive effect, and high cost moves for the proposal at the second stage. As a result the PS produced the smallest number of samples among the MCMC simulations for the same computational cost. Since the tempering factor $\xi$ controls the repulsive effect, it is natural to study the sensitivity of $\xi$ and how it affects the efficiency of the algorithm.

\subsection{Tempering factor $\xi$ in the repulsive proposal} In this section we demonstrate the sensitivity of the tempering factor $\xi$ through a set of examples.

In the MHA, the acceptance probability is
$$\alpha(\theta_i,\varphi_i)=\min\left\{1, \rho(\theta_i,\varphi_i)\right\}$$
where $\rho(\theta_i,\varphi_i)= \displaystyle{\frac{\pi(\varphi_i)q_1(\varphi_i,\theta_i)}{\pi(\theta_i)q_1(\theta_i,\varphi_i)}}$.

The acceptance probability using $\pi^R$ is
$$\alpha^*(\theta_i,\varphi_i)=\min\left\{1, \rho^*(\theta_i,\varphi_i)\right\}$$
where $\rho^*(\theta_i,\varphi_i)= \displaystyle{\frac{\pi^R(\varphi_i)q_1(\varphi_i,\theta_i)}{\pi^R(\theta_i)q_1(\theta_i,\varphi_i)}}$.
\\ 

\begin{figure}[ht!] \begin{center}\setlength{\unitlength}{1cm}
\begin{picture}(7,6)
\put(0,0){\includegraphics[width=7cm, height=6cm]{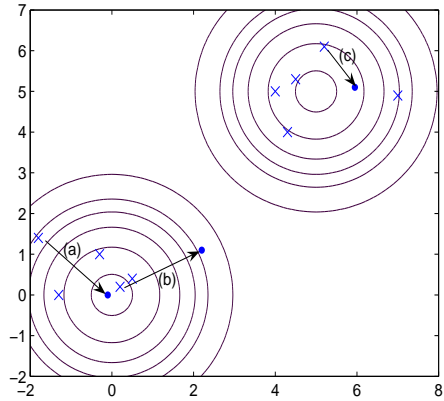}}
%\rput(2.5,2.6) {\small (a)} \rput(4,4) {\small (b)} \rput(2,2.4) {\small (c)}
\end{picture} \end{center} \caption{Ten iid samples (cross) from $\pi$ as in Section 3.1 and three proposals by the normal random walk (dot) from  selected particles which are marked by an arrow.} \label{figure tempering factor}
\end{figure}

The sensitivity of $\xi$ is tested by comparing $\rho$ and $\rho^*$ for different values for $\xi$. Using the same target as in Section 3.1 we consider ten existing particles in which are samples from $\pi$ and random walk proposals depending on selected three particles. This is depicted in Figure \ref{figure tempering factor}; existing particles and proposals are indicated by crosses and dots respectively, and three moves (a)-(c) are represented by arrows.\\

The result is summarized in Table \ref{table tempering factor}, and the acceptance probability is $\rho=\pi(\varphi_i)/\pi(\theta_i)$ and $\rho^*=\pi^R(\varphi_i)/\pi^R(\theta_i)$ due to the use of random walk proposals. It is seen that when $\xi$ is large the $\pi^R$ is dominated by the repulsive term as is the acceptance ratio. In particular for the move (c) the proposal in a high probability region is too close to existing particles and $\rho^*$ decreases rapidly with an increasing $\xi$. In contrast, the proposal in a slightly lower probability region by the move (b) is well separated from existing particles and accepted with a probability of 1 with $\xi=10^{-1}$.\\

\begin{table} $$\begin{array} {|c|c|c|c|c|}
\hline & \rho & \rho^*(\xi=10^{-5}) & \rho^*(\xi=10^{-2})& \rho^*(\xi=10^{-1}) \\ 
\hline {\rm (a)}&13.4237&13.3898&4.3125&1.6399\times 10^{-4}\\
& = \dfrac{0.0792}{0.0059} & =\dfrac{0.0790}{0.0059} & =\dfrac{0.0138}{0.0032} & =\dfrac{2.0719 \times 10^{-9}}{1.2634\times 10^{-5}}\\ \hline
{\rm (b)}&0.0602&0.0603&0.2326&48574\\
& = \dfrac{0.0039}{0.0648} & =\dfrac{0.0039}{0.0647} & =\dfrac{0.0030}{0.0129} & =\dfrac{0.003}{6.1761\times 10^{-9}}\\ \hline
{\rm (c)}&1.1831&1.1831&0.4837&1.4775\times 10^{-4}\\ 
& = \dfrac{0.0504}{0.0426} & =\dfrac{0.0504}{0.0426} & =\dfrac{0.0178}{0.0368} & =\dfrac{1.4775\times 10^{-6}}{0.0100}\\ \hline
\end{array} $$ \caption{Comparison of $\rho = \dfrac{\pi(\varphi_i)}{\pi(\theta_i)}$ and $\rho^*=\dfrac{\pi^R(\varphi_i)}{\pi^R(\theta_i)}$ for different size of repulsive impact.} \label{table tempering factor} \end{table}

In order to observe its effect on the MCMC simulations, we run the MHA with the RP with different values for $\xi$ and estimate the rate of acceptance at the {\it Correction Step} given accepted particles at the {\it Proposal Step}, 
\begin{equation} \label{eqn eta} 
\eta = \dfrac{\mbox{number of accepted particles at {\it Correction Step}}}{\mbox{ number of accepted particles at {\it Proposal Step}}}~.\end{equation} 

\begin{figure}
\begin{center}\setlength{\unitlength}{1cm}  \begin{picture}(8,6)
\put(0,0){\includegraphics[width=8cm, height=6cm]{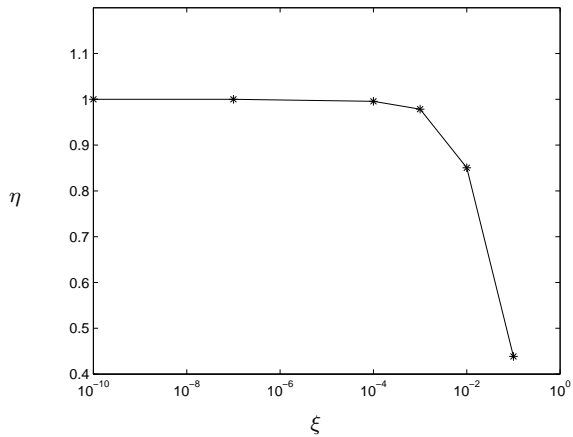}} 
\rput(4,0){\footnotesize {$\xi$}} \rput(0,3){\footnotesize {$\eta$}} 
\end{picture} \end{center} \caption{The rate of acceptance at the Correction stage}\label{figure Temp_RP}
\end{figure}

Figure \ref{figure Temp_RP} shows how $\eta$ varies as the repulsive effect increases based on simulations of 500 iterations. When $\xi<10^{-3}$ most of the proposals that are accepted at the {\it Proposal Step} are accepted at the {\it Correction Step}. As $\xi$ exceeds $10^{-3}$, fewer proposals are accepted and the efficiency of the algorithm decreases rapidly.\\

For efficient sampling, the tempering factor $\xi$ should be tuned, so the proposals are accepted using both $\pi^R$ and $\pi$ at a reasonable rate. In other words, $\pi^R$ should be relatively close to $\pi$ with a value for $\xi$ large enough to give a repulsive effect on the particles.

The tempering factor can be tuned over iterations without changing the stationarity of the chains. For instance a larger value can be used at the beginning of the simulation in a way that induces particles to forget the initial states quickly. 

\subsection{Simulation study : PMC algorithm} 

\subsubsection{Performance study}

The performance comparison of the PMC algorithm and its hybrid approach is carried out in the same manner to that described in Section 3.1, using 100 replicates. The algorithm is run for 500 seconds using 50 particles and the first 100 iterations are ignored. The hyperparameters $\xi=10^{-5}$ and $\nu=0.3$ are used for the repulsive effect.\\

The simulation results based on the 100 repeats are summarized in Table \ref{table PMC}. Here $\tau_{\pi}$ is the integrated autocorrelation time along a series of estimates using the IS, and the $ESS$ estimates differences between the trial distribution based on particles and the target distribution. The $n_b$ denotes the total number of biased estimates during $T$ iterations. The bias is determined by the distance between the estimate of both components using the IS and their expected values, $[\mathbb{E}_{\pi}(\theta_1),\mathbb{E}_{\pi}(\theta_2)]$. If the distance is greater than 3.51, the empirical distribution is asserted to be strongly biased towards one mode over the other mode. The ratio $n_b/T$ indicates the rate of poor exploration and relates to $\tau_{\pi}$.\\

From Table \ref{table PMC} it is seen that the estimate $H$ of both algorithms converge to the expected values reasonably well. It is seen that $\bar{\tau}_{\pi}$ and $\overline{n_b/T}$ of $PMC^{R}$ are reduced, and in particular when $k=2.5$, $\bar{\tau}_{\pi}$ is reduced significantly. This is the most noticeable improvement of fast exploration of particles in the multimodal space among the hybrid algorithms considered here. Moreover there is no sign of the instability of the performance that was a crucial concern shown in the MCMC algorithms.\\

When the scaling parameter $k$ is large so that the $g$ overlap over the probability region, particles move around the space quickly without the repulsive effect. However interestingly the statistical efficiency of the PMC using a wider tailed $g$ is not substantively improved compared to the hybrid approach using a shorter tailed $g$. This means that the statistical performance of this hybrid algorithm is less sensitive to the scaling parameter. Recalling that the performance of the hybrid versions of the MCMC algorithms are sensitive to the proposal variance, this result is useful information in practice.\\

\begin{figure}[ht!] \begin{center}\begin{pspicture}(0,0)(12,4) \psset{linewidth=1cm}
\put(0,0){\includegraphics[width=12cm, height=4cm]{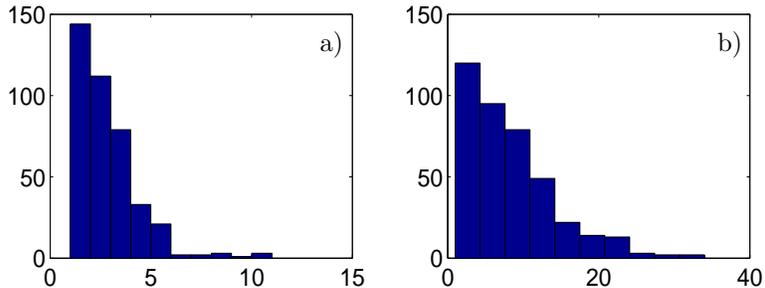}} 
\rput(5.3,3.3){a)} \rput(10.6,3.3){b)}
\end{pspicture} \end{center} \caption{Histogram of the numbers of iterations before the second mode is detected by running the PMC with (a) the repulsive effect, and (b) the PMC for $k=2.5$ during 400 replicates.} \label{figure ModeDetect_pmc}\end{figure}

The mode detecting test was carried out in the same manner as in Section 3.1.2 with resampled particles $(\theta_1,\dots, \theta_N)$ for $k=2.5$. For the PMC, 120 out of 400 repeats detected the mode within the first 5 iterations, with 335 for the PMC with the repulsive effect (Figure \ref{figure ModeDetect_pmc}). This result supports the improvement in the exploration of particles by the use of the repulsive effect.\\

As observed for the MCMC algorithms, adapting the repulsive proposal ($MHA^{RP}$ and $PS$) induces an increase in computational workload per iteration by approximately a factor of 2.5 on average. From this simulation result, we may argue that with 50 particles the gain in statistical efficiency overcomes the drawback of increased demand in computation. However this is not guaranteed since the computational cost increases with increasing $N$.

\begin{table} 
$$ \begin{array} {c|c c c c c c c}
k & & \overline{T} & \bar{H} & \bar{\sigma}^2_H & \bar{\tau}_{\pi}(H) & \overline{n_b/T} & \overline{ESS} \\ \hline 
2.5 & PMC & 35318 & 2.4971 & 1.2727 & 9.6684 & 0.0341 & 8.5958\\
 && (390.7247) & (0.0014) & (0.0027) & (0.0862) & (0.0003) & (0.0035)\\
 & PMC^{R} & 14812 & 2.5003 & 1.1787 & 2.6387 & 0.0198 & 7.8088\\
 && (110.9910) & (0.0013) & (0.0039) & (0.0523) & (0.0002) & (0.0031)\\ \hline
4 & PMC & 40801 & 2.5007 & 1.7669 & 3.0443 & 0.0418 & 5.2150 \\
&& (219.7172) & (0.0009) & (0.0013) & (0.0117) & (0.0001) & (0.0013) \\
& PMC^{R} & 13071 & 2.5011 & 1.7931 & 2.4847 & 0.0408 & 5.9102 \\
&& (50.4404) & (0.0015) & (0.0064) & (0.0649) & (0.0003) & (0.0027) \\ \hline
\end{array} $$ \caption{Comparing of performance of $PMC$ and $PMC^R$ in the two dimensional example. The values in brackets are MSE of measures.} \label{table PMC} \end{table}

\subsubsection{Sensitivity of the repulsive effect}

In order to study the hyperparameters $\xi$ and $\nu$ in the repulsive effect, we compare $\pi$ and $\pi^{2R}$ using $N=50$ particles for different values 
$$\pi^{2R}(\theta_i) \propto \pi(\theta_i) \left((1-\nu)+\nu \prod^N_{j\neq i}e^{-\xi/\pi(\theta_j)\|\theta_i-\theta_j\|^2}\right)~.$$
As an alternative to (\ref{NomCont}) the normalizing constant of $\pi^{2R}$ can be approximated numerically using the finite grid system over the state space. The ratio of the approximate normalizing constants of $\pi$ and $\pi^{2R}$ measures the relative size of holes created over the target distribution and is an indicator of the impact of the repulsive effect.\\

\begin{figure}[ht!] \begin{center}\setlength{\unitlength}{1cm}  \begin{picture}(16,4)
\put(0,0){\includegraphics[width=4cm,height=4cm]{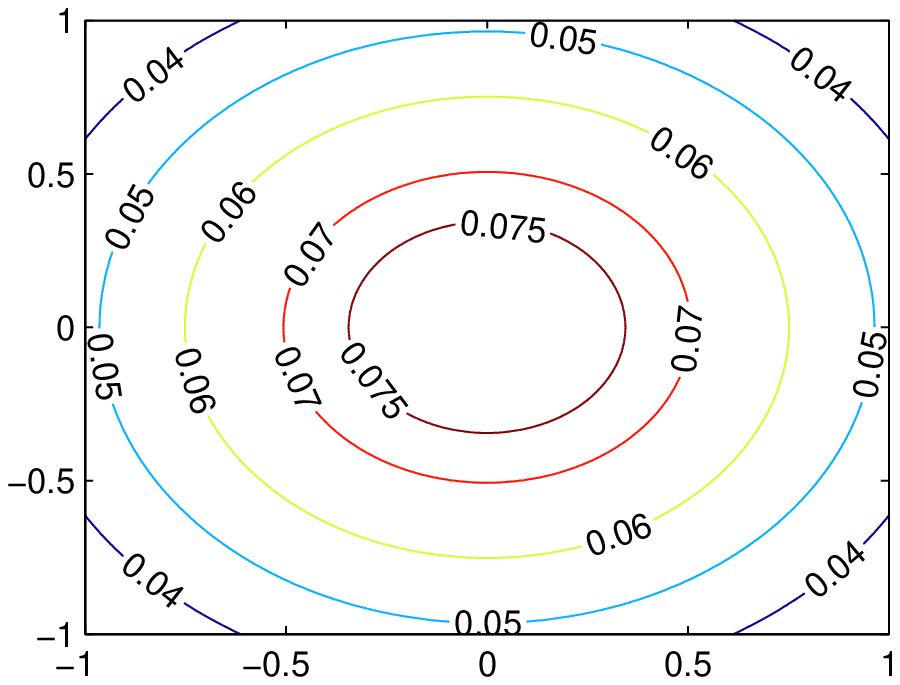}}
\put(4,0){\includegraphics[width=4cm, height=4cm]{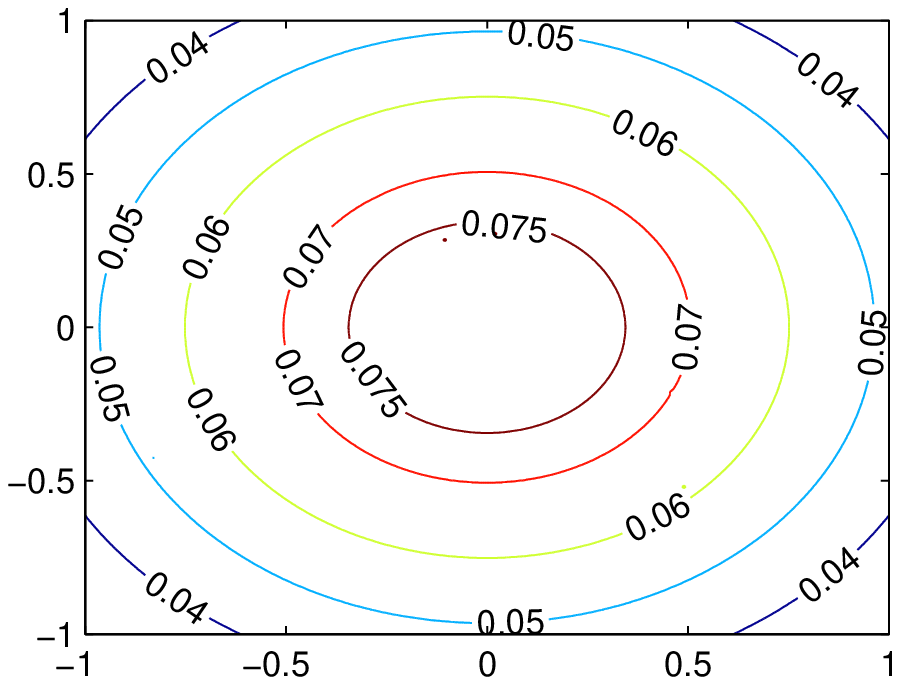}}
\put(8,0){\includegraphics[width=4cm, height=4cm]{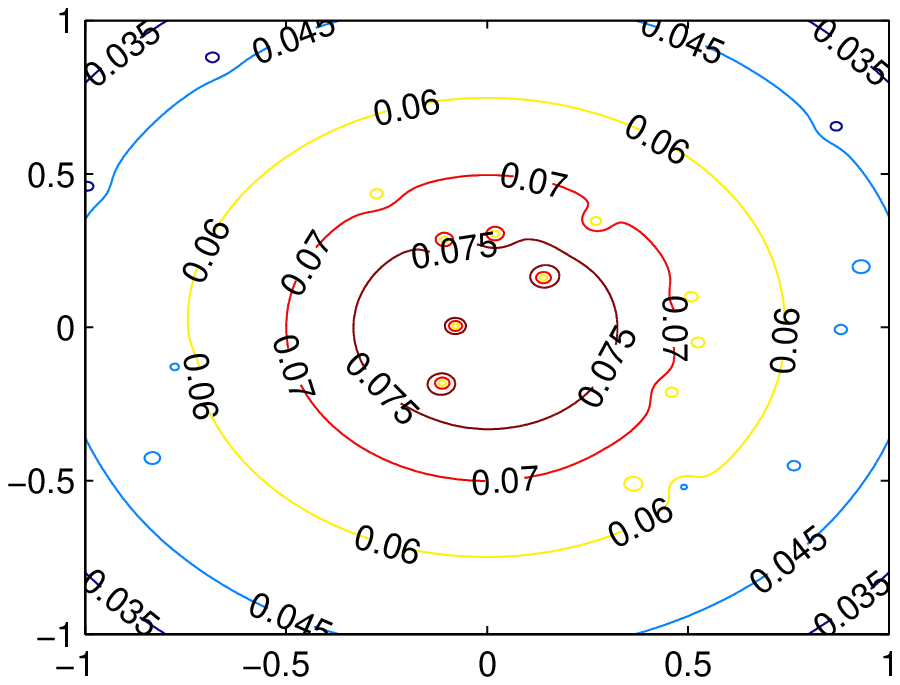}} 
\put(12,0){\includegraphics[width=4cm, height=4cm]{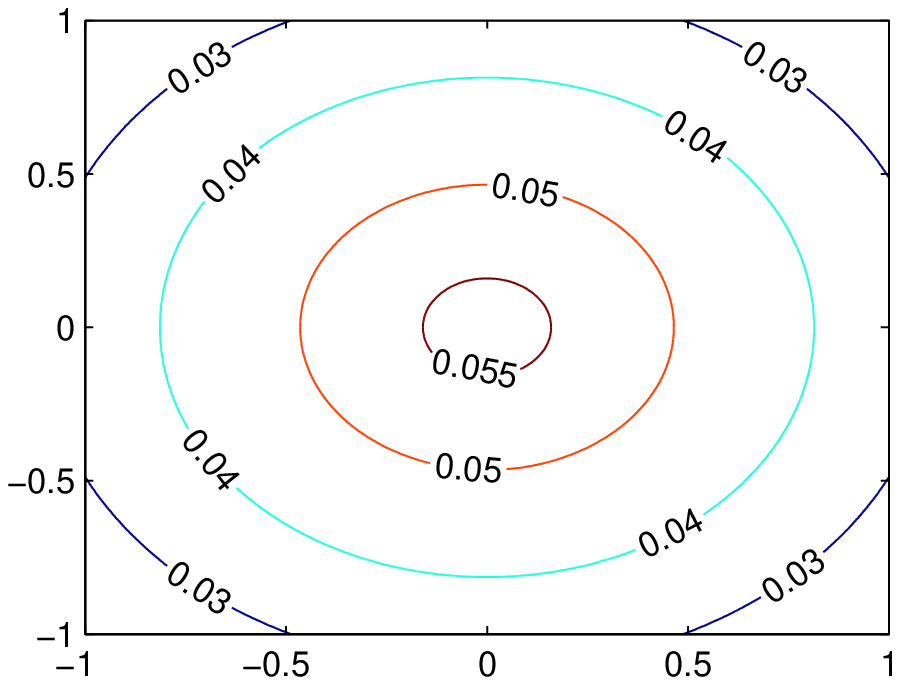}}
\rput(2,-.2){\scriptsize {$\xi=0$}} \rput(6,-0.2){\scriptsize {$\xi=10^{-8}$}}
\rput(10,-.2){\scriptsize {$\xi=10^{-5}$}} \rput(14,-.2){\scriptsize {$\xi=10^{-1}$}}
\end{picture} \end{center}
\caption{The contour plot of $\pi^{2R}$ with $\nu=0.3$ and $\xi=0,10^{-8},10^{-5},10^{-1}$.}\label{figure Contour xi}
\end{figure}

\begin{figure}[ht!] \begin{center}\setlength{\unitlength}{1cm}  \begin{picture}(16,9)
\put(0,.5){\includegraphics[width=8cm, height=4cm]{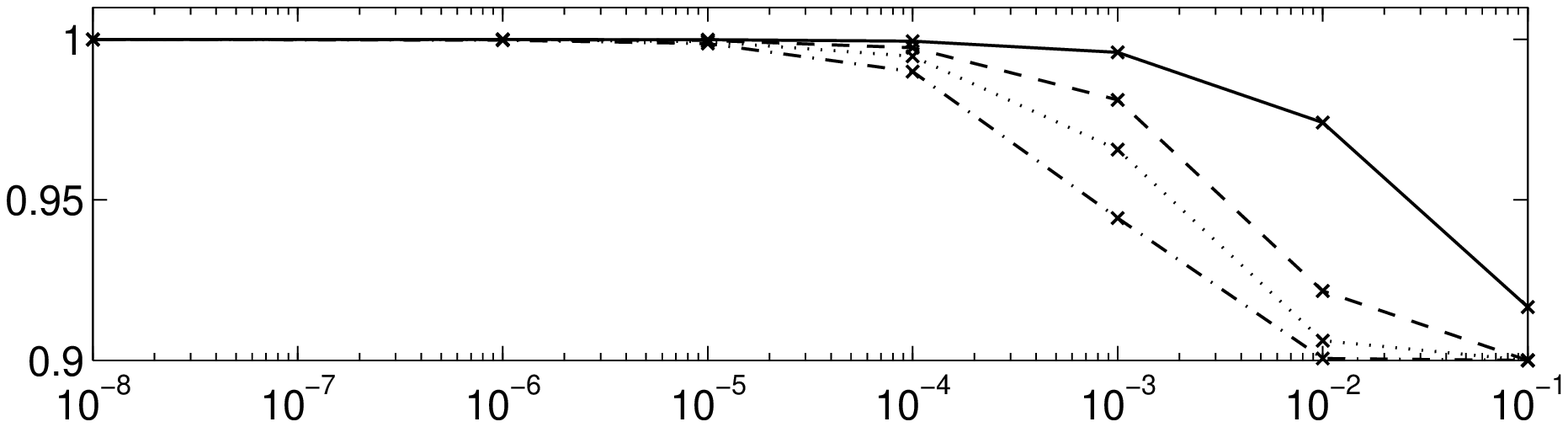}} 
\put(8,.5){\includegraphics[width=8cm, height=4cm]{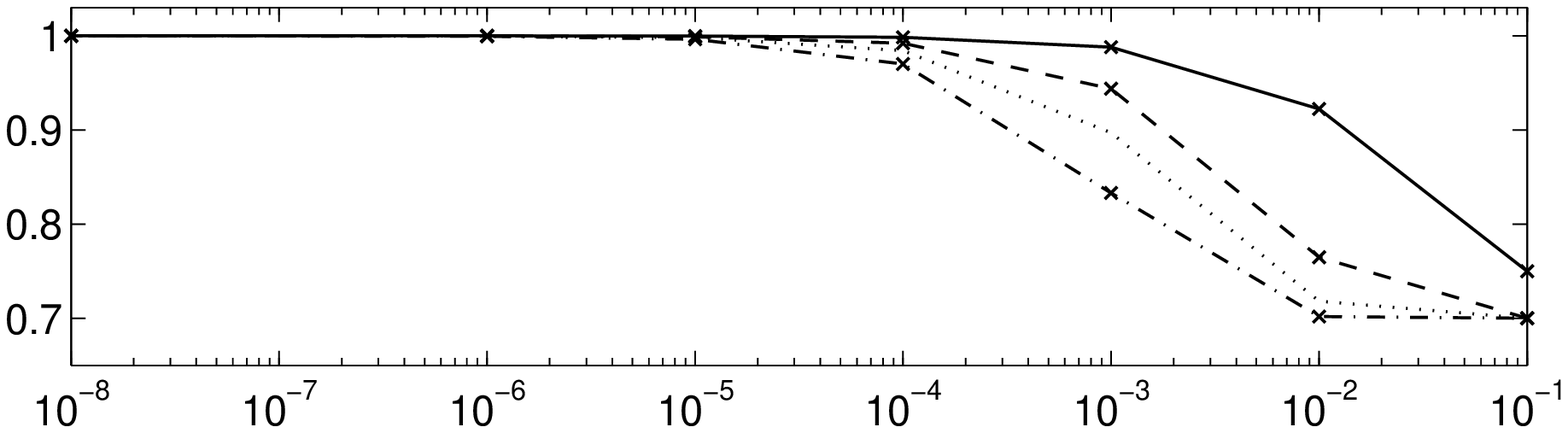}} 
\put(0,5){\includegraphics[width=8cm, height=4cm]{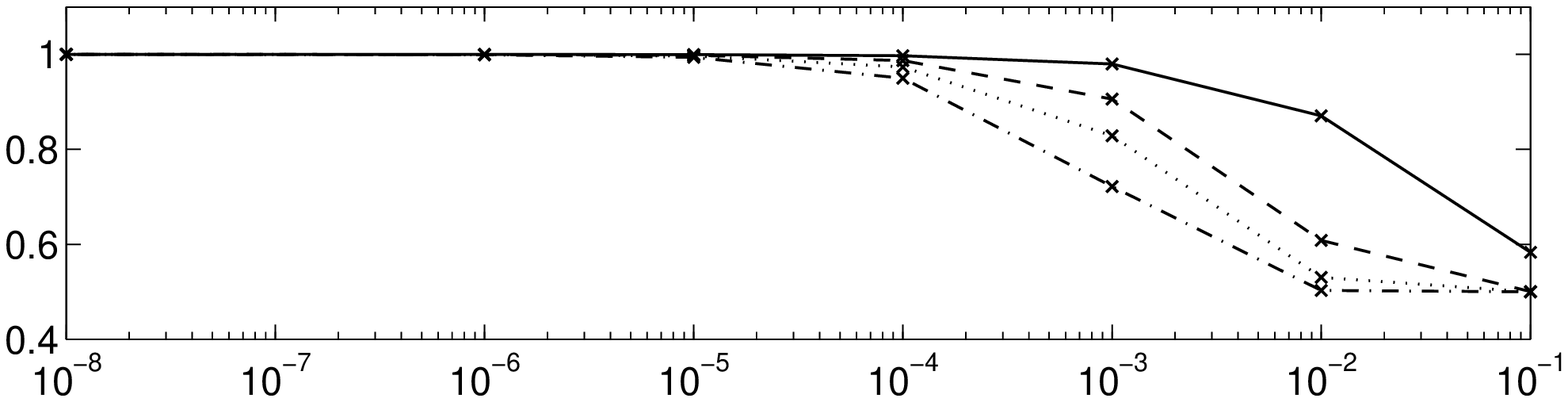}} 
\rput(4,.1){\footnotesize $\nu$}
\rput(12,.1){\footnotesize $\nu$}
\rput(4,4.8){\footnotesize $\nu$}
\rput(0.4,2.5){\footnotesize $C'$} \rput(8.3,2.5){\footnotesize $C'$} \rput(0.4,7){\footnotesize $C'$}
\rput(6.7,3.7){a)} \rput(14.7,3.7){b)} \rput(6.7,8.3){c)}
\end{picture} \end{center}
\caption{The ratio of approximate normalizing constant of $\pi^{2R}$ over $\pi$, $C' \approx \int \pi^{2R}(\theta) d\theta /\int \pi(\theta) d\theta$, for (a) $\nu=0.1$, (b) $\nu= 0.3$, (c) $\nu=0.5$ using $N=10$ (solid line), $N=50$ (dashed line) $N=100$ (dotted line) and $N=200$ (dash-dot line) particles.}\label{figure hyperpara test}
\end{figure}

Figure \ref{figure Contour xi} presents the contour plots of $\pi^{2R}$ using various values for $\xi$. It is seen that the density becomes smooth with no indication of holes when $\xi$ is either very small or very large. For further investigation we estimate the ratio of normalizing constants for various sets of hyperparameters and number of particles (Figure \ref{figure hyperpara test}). As $\xi$ increases, wide holes are overlapped and $\pi^{2R}\approx (1-\nu) \pi$ which has negligible repulsive effect. As $N$ increases the ratio decays to $1-\nu$ quickly with increasing $\xi$. This suggests the use of a smaller $\xi$ when $N$ is large for a significant repulsive effect.\\

The approximation $C'$ as in equation (\ref{NomCont}) is useful in practice where the numerical approximation cannot be evaluated. Its accuracy relates to the number of samples $M$ with respect to the size of holes $\xi$ and $\nu$. For a simple target like our toy example the $M$ can be tuned by adjusting the $C'$ to be close to the ratio of normalizing constant approximations based on the grid system. This process is sensible because the $C'$ can be seen as the ratio of analytically evaluated normalizing constants of $\pi$ and $\pi^{2R}$.\\

In the previous section, the performance study of the $PMC^R$ is undertaken based on simulations using $\xi=10^{-5}$ and $\nu=0.3$. With these values, $\pi^{2R}$ is very similar to $\pi$ and still the repulsive effect is observable. The value of $M$ is chosen to be 50 holes and no additional particles are needed to describe the size of holes.

\section{Application : Aerosol particle size} Aerosols are small particles in suspension in the atmosphere and have both direct and indirect effects on the earth's climate. Aerosol size distributions describe the number of particles observed to have a certain radius, for various size ranges, and are studied to understand the aerosol dynamics in environmental and health modeling.\\

\begin{figure}[ht!] \begin{center}\begin{pspicture}(0,0)(8,5) \psset{linewidth=1cm}
\put(0,0){\includegraphics[width=8cm,height=5cm]{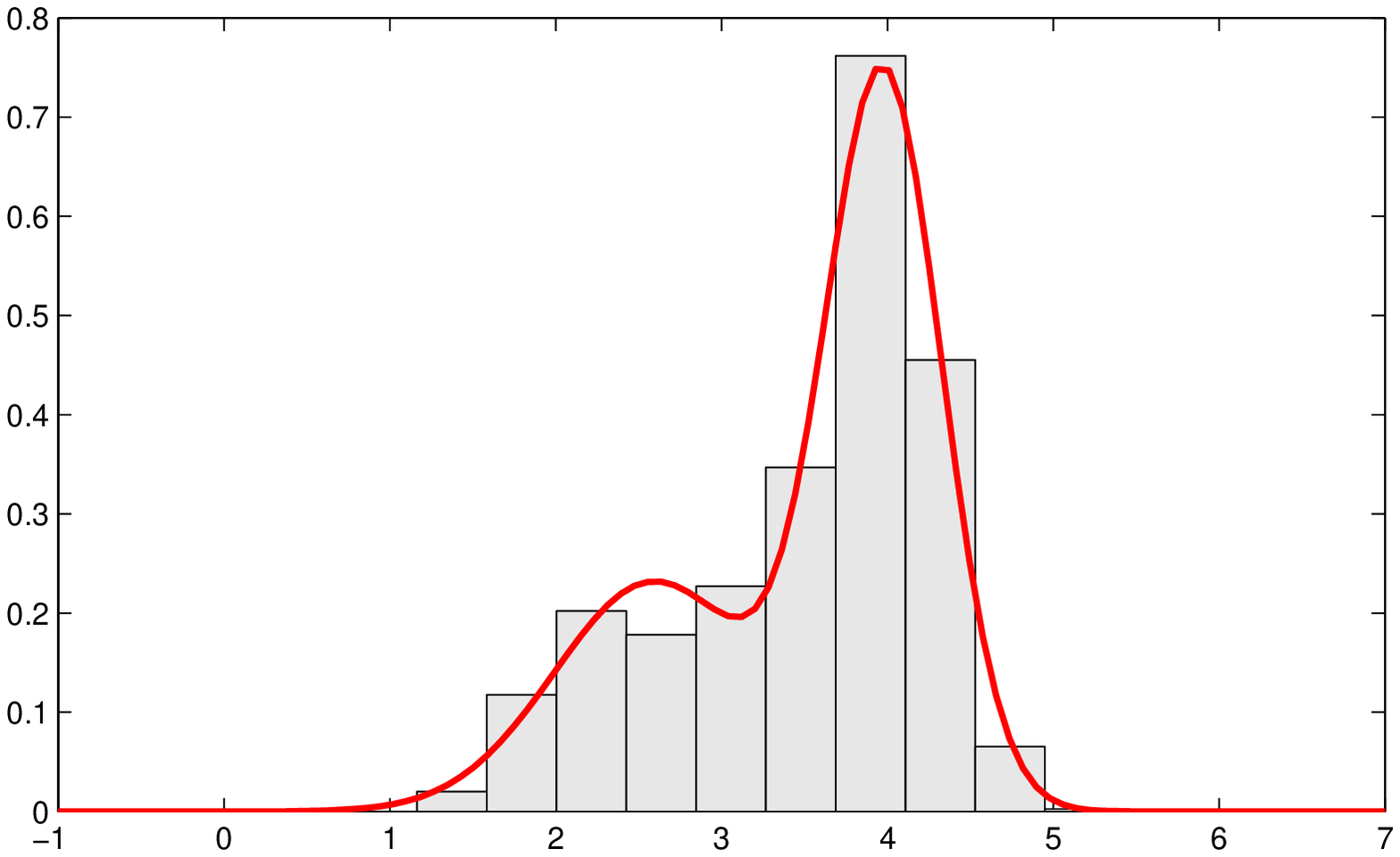}} 
\rput(4,-.2){{\footnotesize Diameter ($nm$)}} \rput{90}(.1,2.5){{\footnotesize Density}}
\end{pspicture} \end{center} \caption{Histogram of the aerosol diameter dataset, along with empirical density via the $MHA$ (solid line) and $PS$ (dotted line).}\label{figure Aerosol}\end{figure}

The data represented in Figure \ref{figure Aerosol} comprise a full day of measurements, taken at ten minutes intervals. The data set was collected at a Boreal forest measurement site at Hyyti\"al\"a, Finland \citep{NilssonKulmala2006}; a random subsample of 2000 observation from the total of 19998 measurements was taken for easier implementation of the algorithms.

The dataset $y=(y_1,\dots,y_{2000})$ is described with a mixture of two normal distributions
$$\lambda\mathcal{N}(\mu_1,\sigma_1^2) + (1-\lambda) \mathcal{N} (\mu_2,\sigma_2^2)$$
where $0<\lambda<1$. The unknown parameters ($\mu_1,\mu_2,\sigma_1,\sigma_2$ and $\lambda$) are estimated using the missing data approach \citep{MarinRobert2007} with relatively vague priors
$$\mu_1,\mu_2 \sim \mathcal{N} (Mean(y),Var(y))~, \hspace{1cm} \sigma_1,\sigma_2\sim \mathcal{G} (2,2) ~, \hspace{1cm} \lambda\sim U[0,1]~. $$

In light of the high dimensionality of the problem we use block updating in which an acceptance probabilities based on the ratio of the full conditional distributions. In addition, considering smaller dimensional components may reduce the sensitivity of the repulsive term in the RP compared to a full dimensional update and allows the use of the geometrical deterministic proposal.\\

The MHA and four hybrid algorithms (the MALA, DRA, DRA with the Langevin proposal and PS) were run using 10 particles for $4200$ seconds. The sets of the parameters ($[\mu_1,\mu_2]$, $[\sigma_1,\sigma_2]$ and $\lambda$) were updated in turn. Details of implementing the algorithms are as follows.
\begin{description}
\item [{\bf MHA and DRA :}] Proposals are generated using the normal random walk with a covariance matrix of $25\times 10^{-4}I_2$ for $[\mu_1,\mu_2]$, $2.25\times 10^{-4}I_2$ for $[\sigma_1,\sigma_2]$ and a variance of $10^{-4}$ for $\lambda$. These updates are used for $DRA^{LP}$ and $PS$ as the first proposal scheme. \\

\item [{\bf MALA and DRA with Langevin proposal :}] At the second stage $[\mu_1,\mu_2]$, $[\sigma_1,\sigma_2]$ and $\lambda$ are updated using the Langevin proposal with $h_{\mu}$, $h_{\sigma}$ and $h_p$ respectively,
$$h_{\mu}=25\times 10^{-4} I_2~, \hspace{.5cm} h_{\sigma}=2.25\times 10^{-4} I_2~, \hspace{.5cm} h_{\lambda}=10^{-4}~.$$

\item [{\bf PS :}] The two dimensional updates $[\mu_1,\mu_2]$ and $[\sigma_1,\sigma_2]$ are implemented according to the algorithm described in Section 2.4 and $\lambda$ is updated as in the MHA, without the delayed rejection mechanism and the RP.\\

For the RP, values for the tempering factor $\xi$ for $\mu$ and $\sigma$ are chosen separately using the acceptance rate at the {\it Correction Step} as in equation (\ref{eqn eta}). Based on Figure \ref{figure Temp_RP_Aerosol} we set $\xi_{\mu}=\xi_{\sigma}=10^{-48}$. Note that these values are quite extreme because they are fixed valued parameters, not balanced scale parameters.

\begin{figure}[ht!] \begin{center}\setlength{\unitlength}{1cm}  \begin{picture}(12,5)
\put(0,0){\includegraphics[width=6cm, height=5cm]{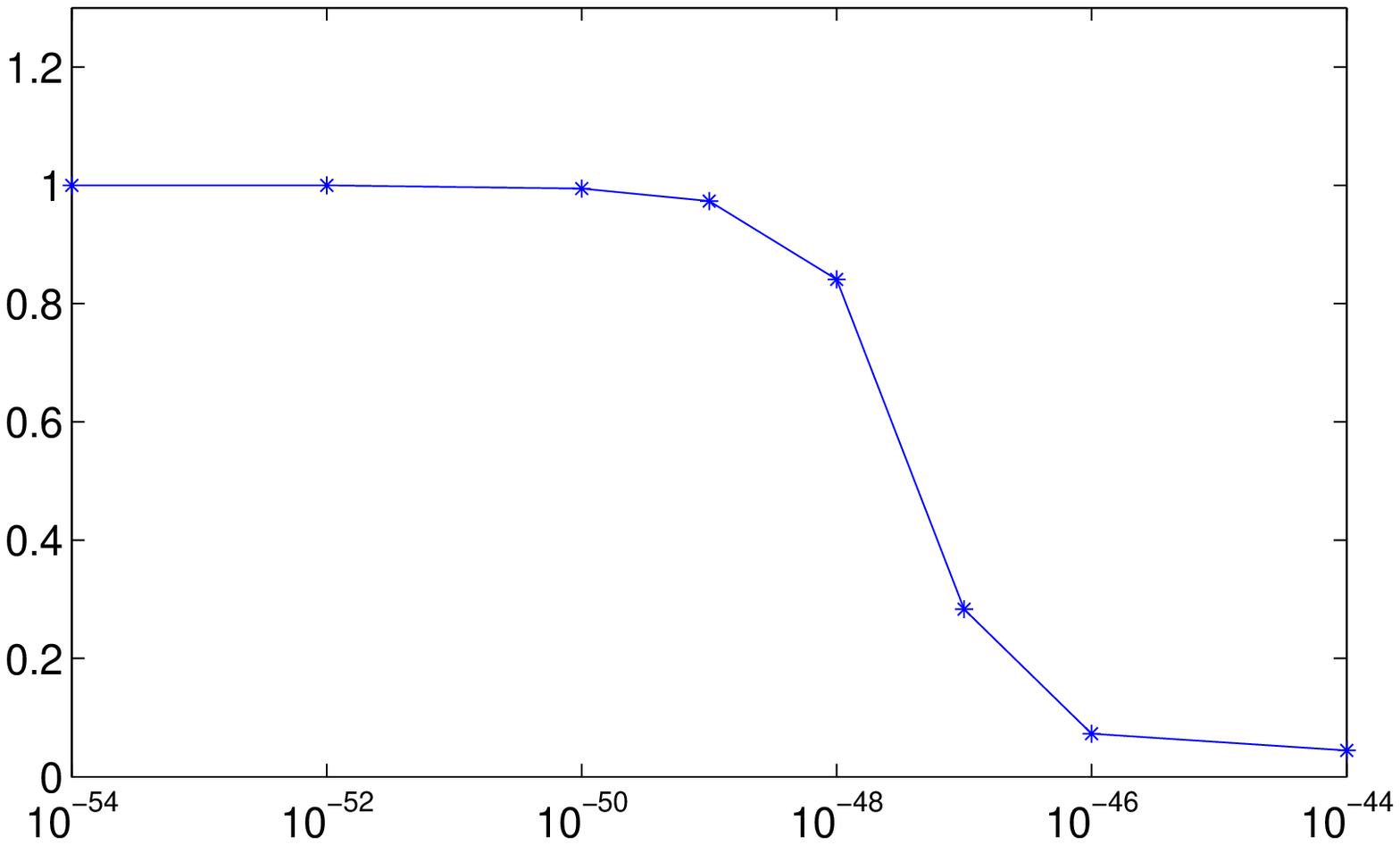}} 
\put(6,0){\includegraphics[width=6cm, height=5cm]{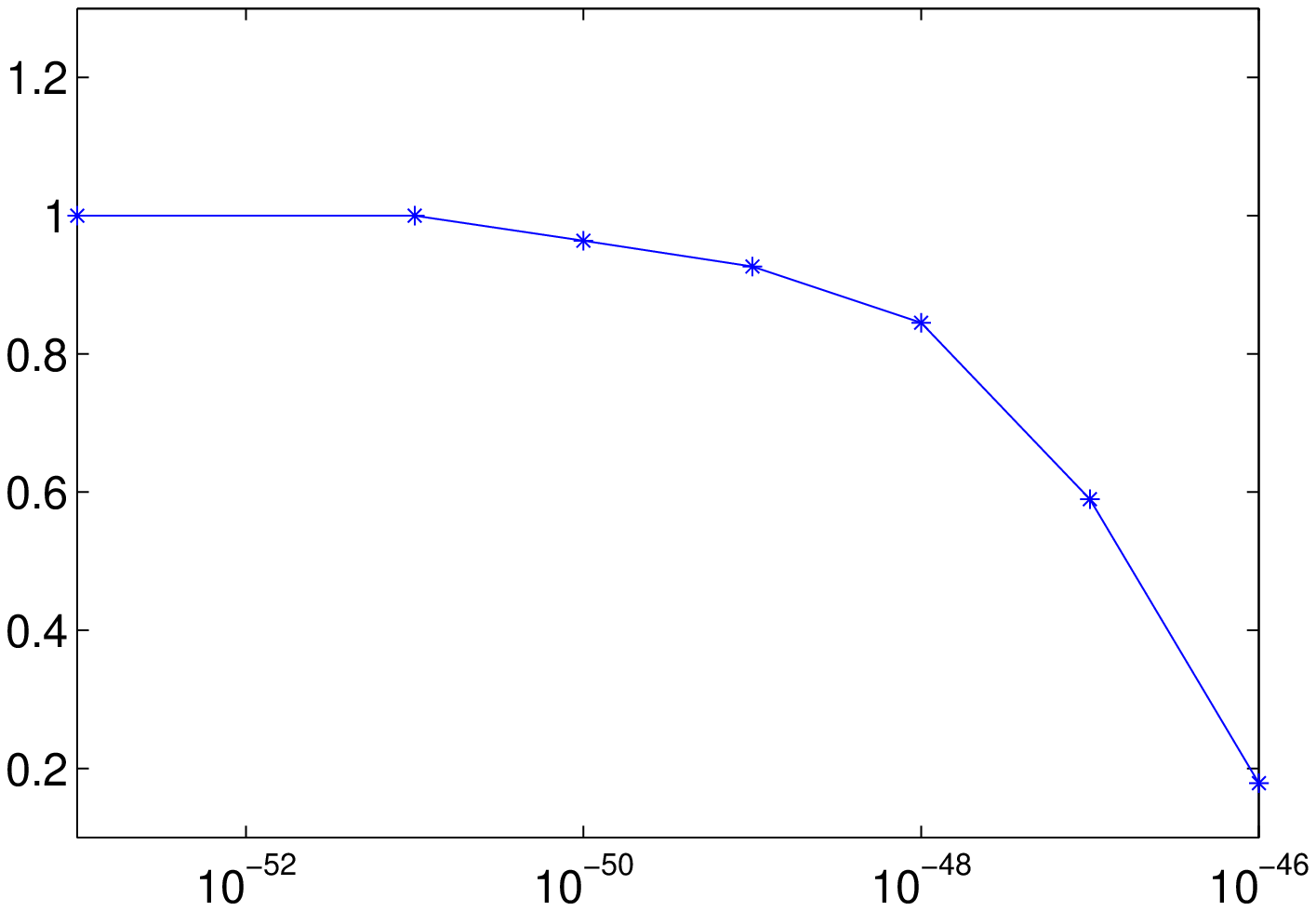}} 
\rput(3,0){\footnotesize $\xi_{\mu}$} \rput(9,0){\footnotesize $\xi_{\sigma}$}
\rput(0,2.5){\footnotesize $\eta$} \rput(6.4,2.5){\footnotesize $\eta$}
\rput(5,4.2){a)} \rput(11,4.2){b)}
\end{picture} \end{center}
\caption{The rate of acceptance at the Correction stage $\eta$ in equation (\ref{eqn eta}) for (a) $[\mu_1,\mu_2]$ and (b) $[\sigma_1,\sigma_2]$ updating.}\label{figure Temp_RP_Aerosol}\end{figure}

\end{description}

\begin{figure}[ht!] \begin{center}
\begin{pspicture}(0,0)(15,12) \psset{linewidth=1cm}
\put(0,8){\includegraphics[width=7cm,height=4cm]{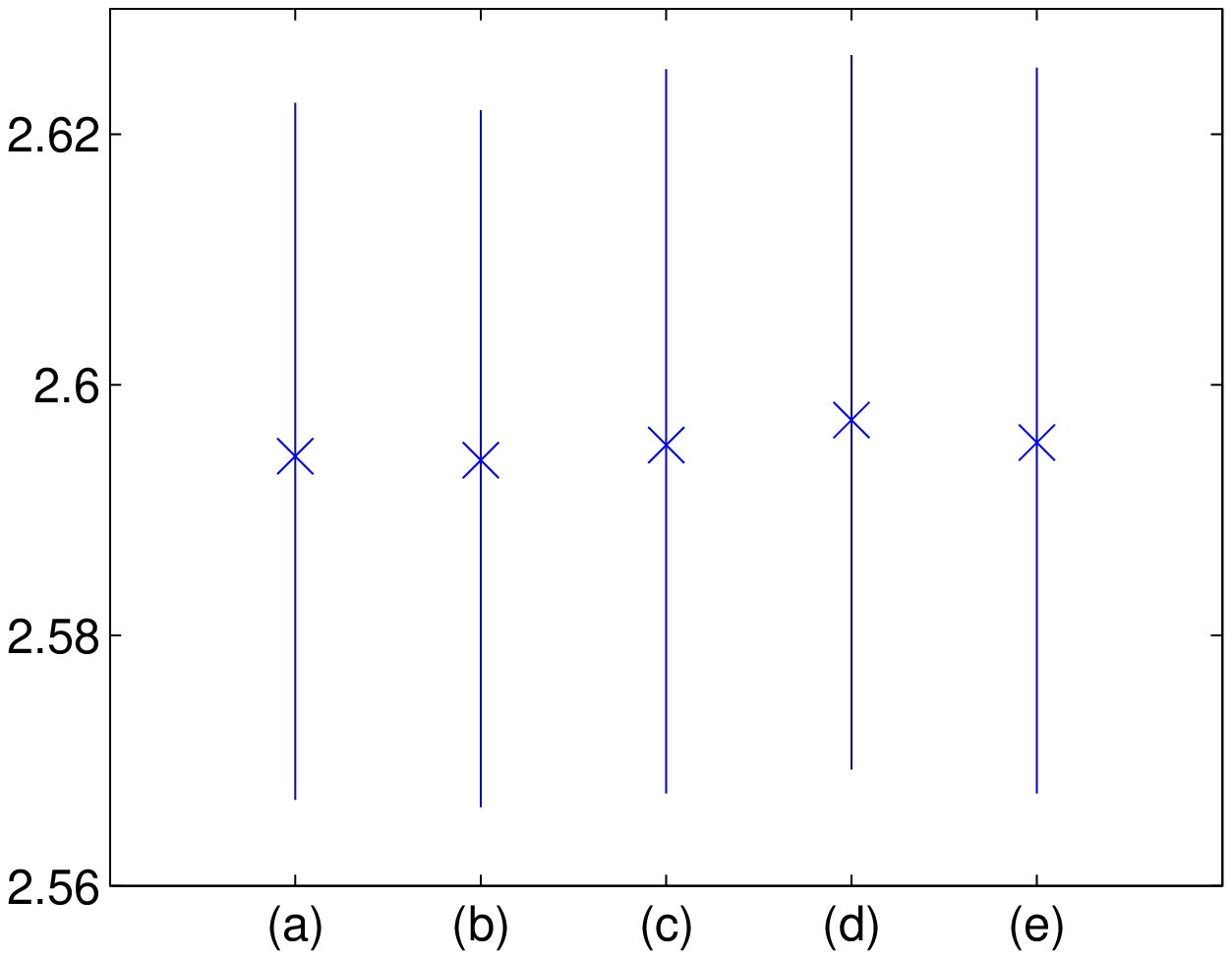}}
\put(8,8){\includegraphics[width=7cm,height=4cm]{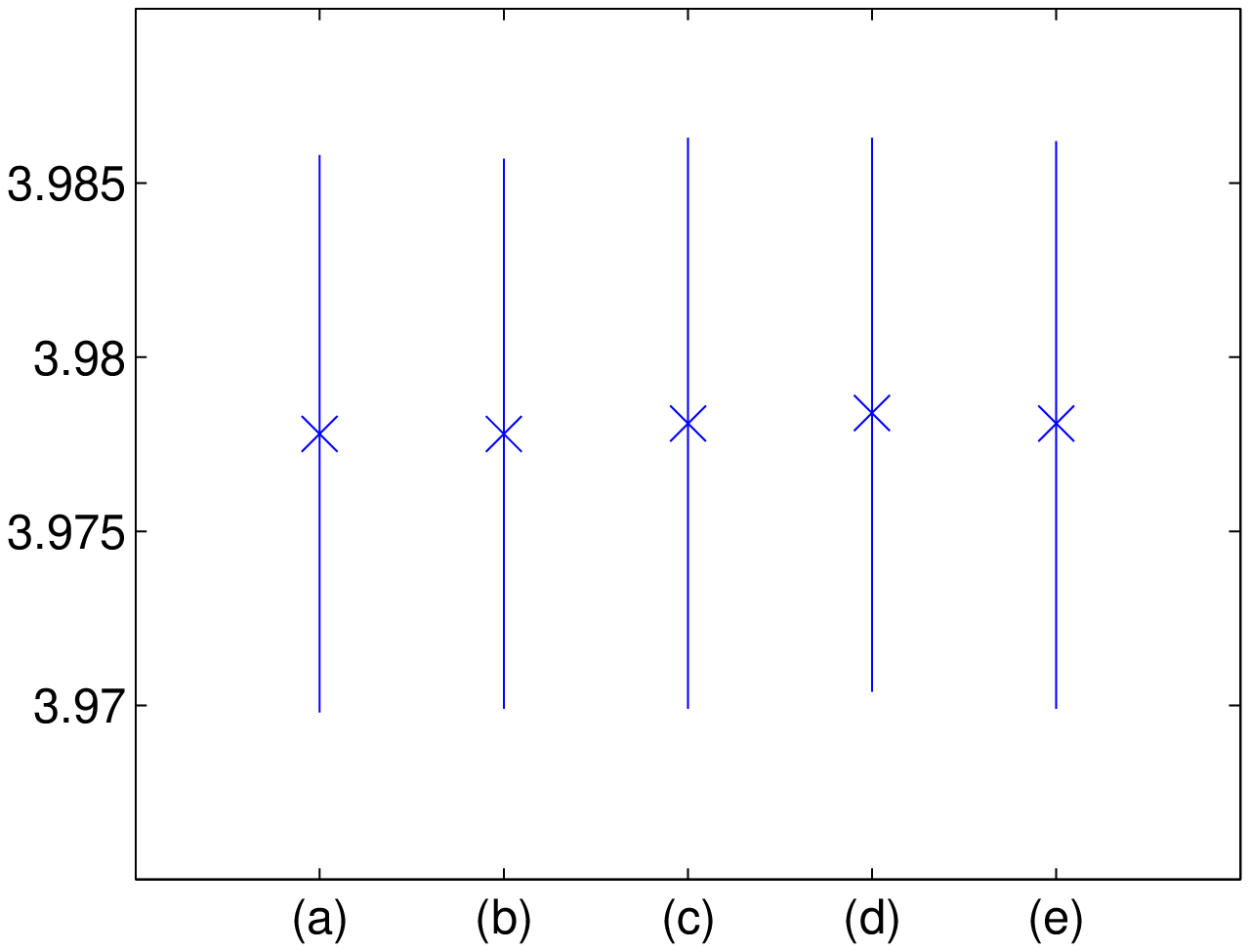}}
\put(0,4){\includegraphics[width=7cm,height=4cm]{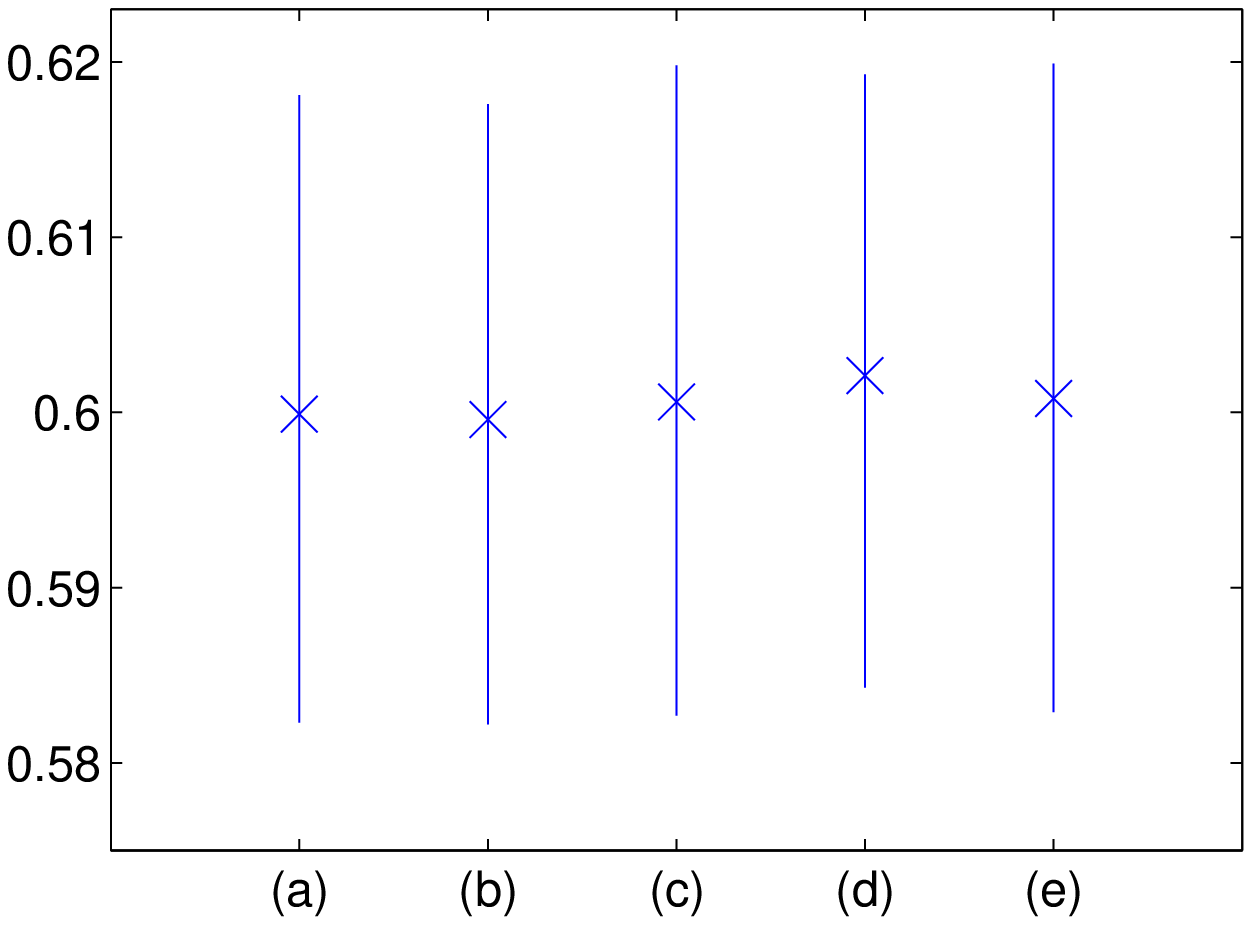}}
\put(8,4){\includegraphics[width=7cm,height=4cm]{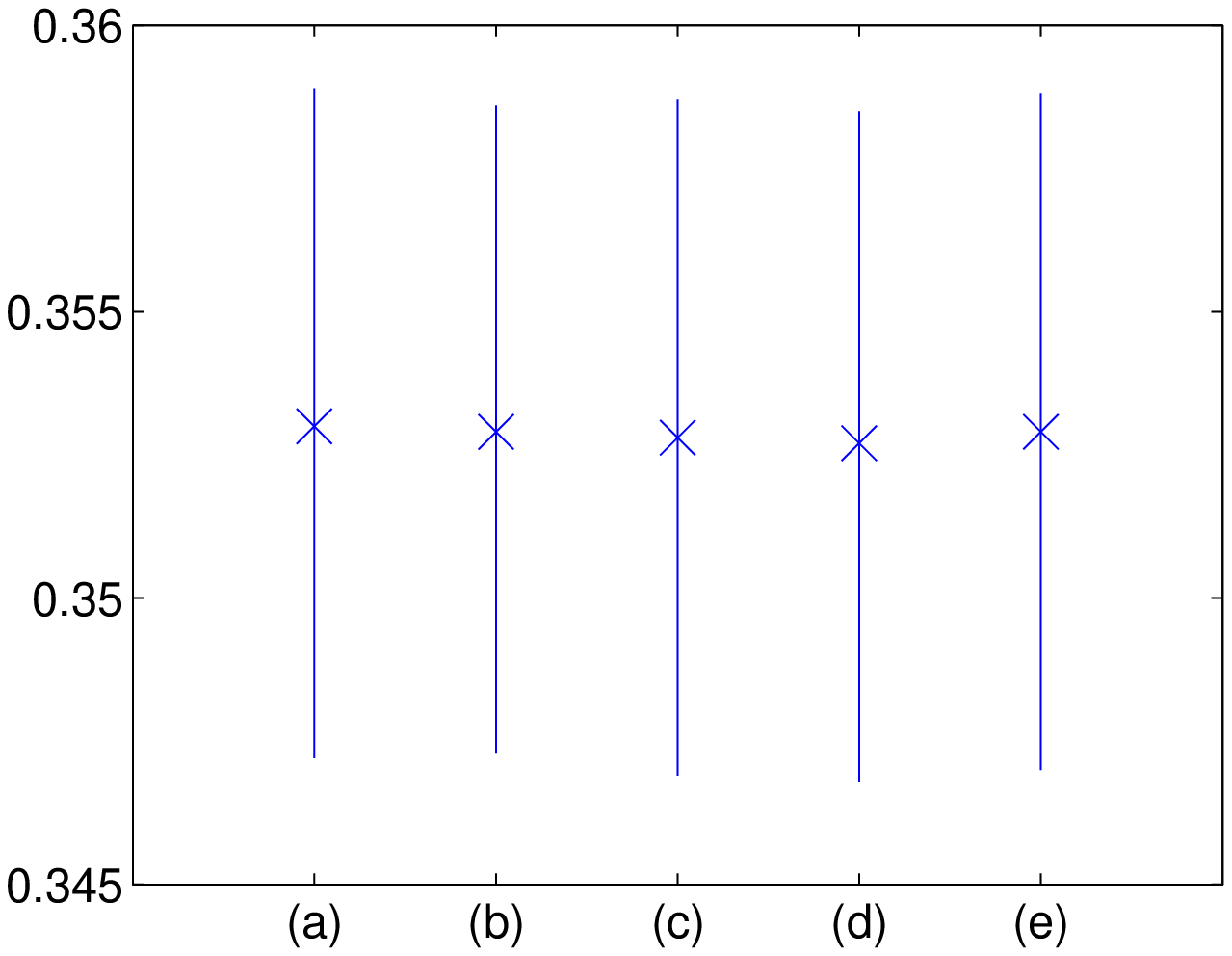}}
\put(0,0){\includegraphics[width=7cm,height=4cm]{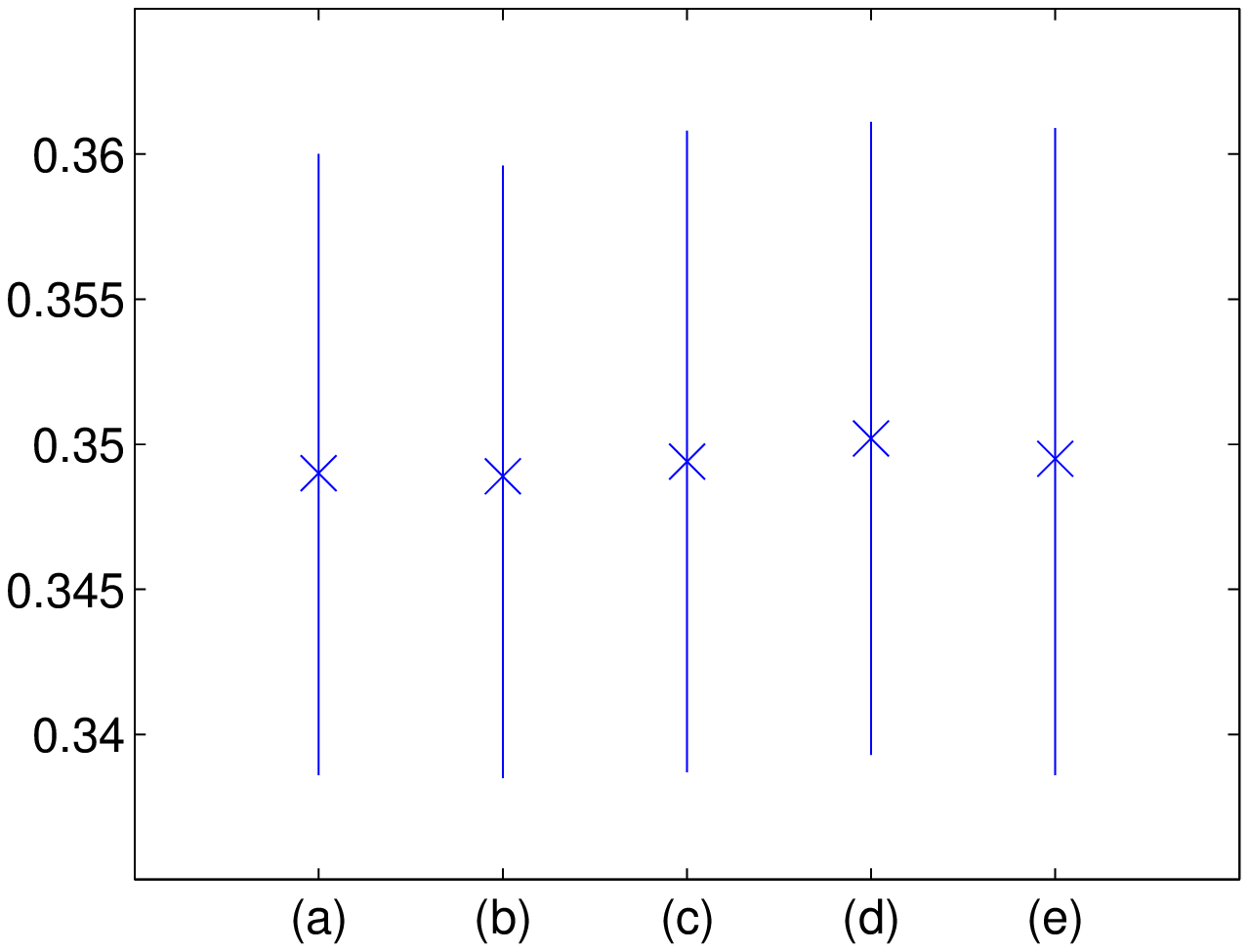}}
\rput{90}(-.1,10){\footnotesize $\mu_1$-values} \rput{90}(7.9,10){\footnotesize $\mu_2$-values}
\rput{90}(-.1,6){\footnotesize $\sigma_1$-values} \rput{90}(7.9,6){\footnotesize $\sigma_2$-values}
\rput{90}(-.1,2){\footnotesize $\lambda$-values}
\end{pspicture} \end{center} \caption{Estimate of parameters (marked as a cross) with the 95\% credible interval of (a) $MALA$, (b) $DRA$, (c) $DRA^{LP}$, (d) $PS$, and (e) $MHA$ (marked as a solid line).}\label{Para Aerosol}\end{figure}

\begin{table} $$ \begin{array} {c|c c c}
\hline & T & \bar{\tau}_{\pi} & ESS \\ \hline 
MALA & 46071 & 65.6773 & 686.2 \\
DRA^{LP} & 87613 & 114.1824 & 758.5 \\ 
DRA & 125020 & 130.7077 & 948.8 \\
PS & 20775 & 127.7793 & 154.8 \\
MHA & 203118 & 185.4259 & 1090.0 \\ \hline
%MHA^s & 636848 & 164.4192 & 3867.2 \\
\end{array} $$ \caption {Performance of selected hybrid algorithms ($MALA$, $DRA^{LP}$, $DRA$ and $PS$) and the MHA} \label{table aerosol} \end{table}

\begin{figure}[ht!] \begin{center}\setlength{\unitlength}{1cm}  \begin{picture}(20,8)
\put(-2,0){\includegraphics[width=20cm,height=8cm]{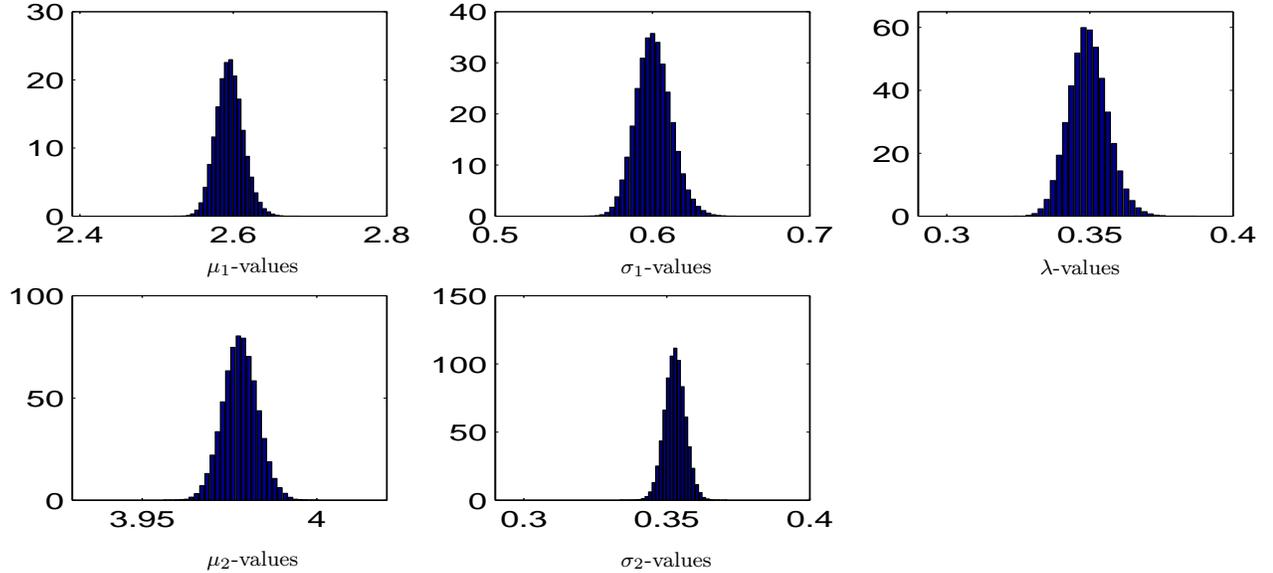}}
\rput(3,4){\footnotesize $\mu_1$-values} 
\rput(3,0.1){\footnotesize $\mu_2$-values}
\rput(8.5,4){\footnotesize $\sigma_1$-values} 
\rput(8.5,.1){\footnotesize $\sigma_2$-values}
\rput(14,4){\footnotesize $\lambda$-values}  
\end{picture} \end{center} \caption{Histogram of parameters ($\mu_1, \sigma_1,\lambda,\mu_1\sigma_2$) of $MHA$.}\label{Fig:Hist_Aerosol}\end{figure}

The results of the MCMC simulations are summarized in Table \ref{table aerosol} and Figures \ref{Para Aerosol} and \ref{Fig:Hist_Aerosol}. Since the two modes are fairly clearly separated, label switching was not observed for in any of the simulations \citep{Geweke2007}. Overall the estimates of the five parameters are similar and the 95\% credible interval of parameters obtained from all algorithms overlap each other. The $\bar{\tau}_{\pi}$ denotes the average of $\tau_{\pi}$'s of all parameters and $ESS=(T-T_0)/\bar{\tau}_{\pi}$ ($T_0=1000$).\\

In general, the features of the hybrid algorithms shown from this simulation are similar to those observed using the toy example in Sections 3.1 and 3.2. The PS is an expensive algorithm to compute and the MHA demands the least CPU time per iteration. The Langevin proposal was very effective in reducing $\tau_f$ and the DRA did not improve the mixing as it did in the toy example. This is supported by the observation that the two modes are well separated and the empirical posteriors of each parameter have a single mode as seen in Figure \ref{Para Aerosol}. By the property of the Langevin diffusion, particles are pushed back to the mode and consequently can explore a single mode more quickly.

\section{Discussion} 

In this paper we have considered the problem of designing hybrid algorithms through studying selected hybrid methods, the DRA with three types of move for the second-stage proposal, the MALA, the PS, and the PMC. Using a two dimensional example, each algorithm was evaluated and the relative contributions of individual components to the overall performance of the hybrid algorithm were estimated. The performance was defined by the accuracy of estimation, the efficiency of the proposal move, the demand in computation, the rate of mixing of the chains and the ability of the algorithm to detect modes in a special setup. Algorithms were also applied a real problem of describing an aerosol particle size distribution. We observed that the results of this analysis coincided with those obtained using a toy example.\\

Based on these simulation studies, we subjectively rated the performance of the hybrid MCMC algorithms relative to the MHA, and the PMC adapting the repulsive effect relative to the PMC. The results are presented in Table \ref{table discussion}. Following the three criteria for a efficient algorithm defined in Section 1, we considered the efficiency of the proposal move (EPM), the correlation reduction of a chain (CR), the rate of convergence (RC), the cost effectiveness (CE), the simplicity of programming (SP), the flexibility of hyperparameters (FH), the consistency of performance (CP) and the preference between a single mode and a multimodal problem (Mode). As the algorithm improves with respect to the criterion, the rating increases positively, with zero indicating that there is no substantive difference to either the MHA or the PMC. Among the MCMC algorithms it is seen that $DRA$ and $DRA^{LP}$ are sensible algorithms to use for multimodal problems and the MALA for unimodal problems. For the hybrid versions via the adaptation of the repulsive effect the overall efficiency based on criteria is improved with the PMC relative to the MHA. 

\begin{table} \centering \begin{tabular}{l|c|c|c|c|c|c|c|c} \hline
Algorithm & \multicolumn{3}{c|}{Statistical Eff.} & \multicolumn{2}{c|}{Computation} & \multicolumn{3}{c}{Applicability} \\ \hline
& EPM & CR & RC & CE & SP & FH & CP & Mode \\ \hline
$MALA$ & 1 & 1 & 1 & -2 & -1 & 0 & -1 & Single \\
$MHA^{RP}$ & 0 & 1 & 0 & -2 & -1 & -1 & -2 & Both \\ 
$DRA$ & 1 & 1 & 0& -1 & -1 & 0 & 0 & Both \\
$DRA^{LP}$ & 2 & 1 & 0 & -2 & -1 & 0 & 0 & Both \\
$DRA^{Pinball}$ & 2 & 1 & 0 & -2 & -1 & 0 & 0 & Both\\
$PS$ & 2 & 1 & 0& -3 & -2 & -1 & -2 & Both \\ 
$PMC^{R}$ & - & 2 & 0 & -1 & -1 & -1 & 0 & Both \\ 
\hline
\end{tabular} \caption{The relative performance rating of the hybrid algorithms to the MHA and the PMC.} \label{table discussion} \end{table}

We identified the following overall issues to be considered in designing hybrid algorithms.

\begin{description}
\item [(i)] Combining features of individual algorithms may lead to complicated characteristics in a hybrid algorithm. For instance, it was seen that the performance of some algorithms was sensitive with respect to the value of the tuning parameters such as $s$, $h$, and $\xi$, and this also applied to the hybrid algorithms. For the MHA with the repulsive effect the improvements from the statistical perspective were sensitive to both the size of the normal random walk (a well known property of the MHA) and the tempering factor. Moreover, some inconsistency of estimates throughout replicates occurred, albeit rarely.

\item [(ii)] Each individual algorithm may have a strong individual advantage with respect to a particular performance criterion, but this does not guarantee that the hybrid method will enjoy a joint benefit of these strategies. The contribution of some components may be insignificant, and certain components may dominate the character of the hybrid method. This can be seen with the PS as the DRA most strongly contributes to statistical efficiency.

\item [(iii)] From the perspectives of applicability and implementation, the combination of algorithms may add complexity in setup, programming and computational expense. These phenomena are easily observed as all hybrid approaches considered in this paper were computationally demanding, although the magnitude of the demand differed with the types of move and techniques. In practice it is important to be aware of these issues to optimize the performance in real time so that the improvement in one criterion does not become negligible due to the drawback in computation. This was observed, for example, with the $DRA^{LP}$.

\end{description}
These general considerations in building an efficient algorithm are easily applied to existing hybrid methods. The adaptive MCMC algorithm is another good example. Its motivation is to automate and improve the tuning of the algorithm by learning from the history of the chain itself. It can be easily coded and is statistically efficient without a strenuous increase in the computational expense. \citet{RobertsRosenthal2007} and \citet{Rosenthal2008} demonstrated that an algorithm can learn and approach an optimal algorithm via automatic tuning scaling parameters to optimal values. After a sufficiently long adaptation period it will converge much faster than a non-adapted algorithm. However this fast convergence feature may increase the risk that chains may converge to the wrong values \citep{RobertCasella2004,Rosenthal2008}. Moreover, since proposals are accepted using the history of the chain the sampler is no longer Markovian and standard convergence techniques cannot be used. The adaptive MCMC algorithms converge only if the adaptations are done at regeneration times \citep{GilksRobertsSahu1998,BrockwellKadane2005} or under certain technical types of the adaptation procedure \citep{HaarioSaksmanTamminen2001, AndrieuMoulines2006,  AtchadeRosenthal2005, RobertsRosenthal2007}. Overall the adaptive MCMC algorithm can be efficient from the statistical perspective and cost effective only when it is handled with care. In other words, these advantages come with a reduction in robust reliability which may affect its applicability.\\

In this paper, hybrid algorithms comprising components running in parallel were considered. It would also be of interest to consider and compare in an analogous fashion hybrid algorithms comprising components that are run in series. Without compromising the stationarity of the chain, a set of components similar to those described here could be implemented either in turn or randomly, with or without cues from the existing chain or estimated distribution. The resultant algorithms would have diverse characteristics related to the selected components as well as the sequence in which they were run. As considered here, parallels with current adaptive MCMC algorithms could also be drawn.

\section*{Acknowledgements}
This work was supported by the Australian Research Council Center for Dynamic Systems and Control. 
%\end{acknowledgements}

\newpage \raggedright
\nocite{*}
\bibliographystyle{plainnat}
\addcontentsline{toc}{chapter}{Bibliography}
\bibliography{reference}

\end{document}